\documentclass[fleqn,usenatbib]{mnras}

\usepackage[T1]{fontenc}
\usepackage{ae,aecompl}
\usepackage{multicol}

\usepackage{enumerate}
\usepackage[normalem]{ulem}
\usepackage[usenames]{color}

\usepackage{graphicx}	
\usepackage{amsmath}	
\usepackage{amssymb}	

\usepackage{newtxtext,newtxmath}

\newcommand{\athena}{{\sevensize ATHENA++}}

\newcommand{\mesa}{{\sevensize MESA}}

\newcommand{\be}{\begin{eqnarray}}
\newcommand{\ee}{\end{eqnarray}}
\newcommand{\rsun}{\ensuremath{\textrm{R}_{\odot}}}

\DeclareMathOperator\arctanh{arctanh}

\newcommand{\grad}{\ensuremath{\boldsymbol{\nabla}}}
\newcommand{\vel}{\ensuremath{\boldsymbol{v}}}

\newcommand{\ddt}[1]{\ensuremath{\frac{\partial #1}{\partial t}}}

\newcommand{\ra}{\ensuremath{R_{A}}}
\newcommand{\rs}{\ensuremath{R_{s}}}
\newcommand{\fdf}{\ensuremath{F_{\rm DF}}}
\newcommand{\smax}{\ensuremath{s_{\rm max}}}
\newcommand{\smin}{\ensuremath{s_{\rm min}}}
\newcommand{\mach}{\ensuremath{\mathcal{M}}}

\newcommand{\meshr}{\ensuremath{\boldsymbol{r}}}

\newcommand{\lp}[1]{\textrm{\color{red} #1}}

\renewcommand{\sout}[1]{}
\renewcommand{\lp}[1]{#1}

\title[Flow Morphology of a Supersonic Gravitating Sphere]{Flow Morphology of a Supersonic Gravitating Sphere}

\author[]{ Logan J. Prust$^1$\thanks{LJP: ljprust@kitp.ucsb.edu}, Lars Bildsten$^{1,2}$
\\
$^1$ Kavli Institute for Theoretical Physics, University of California, Santa Barbara, CA 93106, USA\\
$^2$ Department of Physics, University of California, Santa Barbara, CA 93106, USA
}

\date{Accepted XXX. Received YYY; in original form ZZZ}

\pubyear{2023}

\begin{document}
\label{firstpage}
\pagerange{\pageref{firstpage}--\pageref{lastpage}}
\maketitle

\begin{abstract}
Stars and planets move supersonically in a gaseous medium during planetary engulfment, stellar interactions and within protoplanetary disks. For a nearly uniform medium, the relevant parameters are the Mach number and the size of the body, $R$, relative to its accretion radius, $R_A$. Over many decades, numerical and analytical work has characterized the flow, the drag on the body and the possible suite of instabilities. Only a limited amount of work has treated the stellar boundary as it is in many of these astrophysical settings, a hard sphere at $R$. Thus we present new 3-D \athena\ hydrodynamic calculations for a large range of parameters. For $R_A\ll R$, the results are as expected for pure hydrodynamics with minimal impact from gravity, which we verify by comparing to experimental wind tunnel data in air. When $R_A\approx R$, a hydrostatically-supported separation bubble forms behind the gravitating body, exerting significant pressure on the sphere and driving a recompression shock which intersects with the bow shock. For $R_A\gg R$, the bubble transitions into an isentropic, spherically-symmetric halo, as seen in earlier works. These two distinct regimes of flow morphology may be treated separately in terms of their shock stand-off distance and drag coefficients. Most importantly for astrophysical applications, we propose a new formula for the dynamical friction  which depends on the ratio of the shock stand-off distance to $R_A$. That exploration also reveals the minimum size of the simulation domain needed to accurately capture the deflection of incoming streamlines due to gravity.
\end{abstract}




\section{Introduction} \label{sec:intro}

Supersonic motion of a gravitating body within a gaseous medium arises in many areas of astrophysics, and the shock waves which adorn such objects facilitate both their gravitational and hydrodynamic interactions with the gas. These problems vary in the physics responsible for the generation of the bow shock, such as wind-wind interactions \citep{2014MNRAS.444.2754M,2016A&A...586A.111S,2018MNRAS.477.2431T} or MHD shocks from magnetospheres \citep{2021JGRA..12629104K}. Bow shocks may also be generated through purely hydrodynamical processes, as in the supersonic motion of an accretor. Often, the gravitating body can be modeled as a point source \citep{1971MNRAS.154..141H}, as an accretion boundary \citep{1985MNRAS.217..367S,1994ApJ...427..342R,1994ApJ...427..351R,2012ApJ...752...30B,2019MNRAS.488.5162X} or with a Plummer potential \citep{2001MNRAS.322...67S,2009ApJ...703.1278K,2021arXiv210315848M}. In any case, the body feels a drag force due to dynamical friction, which has been analytically modeled for both subsonic and supersonic motion \citep{1999ApJ...513..252O}. Accreting bodies feel an additional drag equal to the total momentum of the accreted gas.

Also of interest is the case in which the gravitating body has a rigid surface at a fixed radius (or cannot effectively accrete), which has received relatively little study. Rather than an accretion drag, such an object feels a net force due to the non-uniform pressure distribution over its surface. \citet{1980Ap&SS..67..427S} computed the flow structure around a hypersonic rigid sphere, though their analysis focused on isodensity contours in the far field. The dearth of studies on rigid spheres was ameliorated by \citet[][hereinafter T16]{thun}, who carried out a thorough investigation of shock stand-off distance and dynamical friction on such objects in homogeneous media. T16 studied supersonic bodies with Mach numbers $\mach\geq2$ and proposed a new formula for dynamical friction which was demonstrated to perform well for objects with large accretion radii $\ra$ relative to their physical radii $R$. Here the accretion radius is defined as 
\be
\ra = 2 G m/v_{\infty}^{2},
\ee
where $m$ and $v_{\infty}$ are the mass and velocity of the body. Though the gas accumulated at the surface of the sphere is too optically thick to radiatively cool, preventing actual accretion onto the surface, $\ra$ is nevertheless useful as a metric of the strength of gravity relative to inertial forces.



Flow around a rigid gravitating sphere is primarily relevant to problems in binary stellar interactions or planet-star interactions. One example is the interaction of a star with supernova ejecta from its companion, which was studied by \citet{2019ApJ...887...68B} as a possible formation mechanism for runaway stars in unusual locations on the Hertzsprung-Russell diagram. Another is the migration and orbital evolution of planetesimals in a protoplanetary disk, which can take place due to aerodynamic drag and dynamical friction as well as other effects such as tides and Lindblad resonances \citep{2015ApJ...811...54G}. The atmosphere of a protoplanet is also affected by the presence of a bow shock \citep{2020ApJ...899...54M}, though protoplanets generally do not develop such shocks unless they are on eccentric orbits \citep{2021ApJ...915..113B}. The use of a rigid surface to model the body in these scenarios is well-motivated by the large entropy contrast between the stellar or planetary material comprising the body and the shocked gas surrounding it.

A prime example is the engulfment of a planet by its host star. As a star ascends the red giant branch (RGB) or asymptotic giant branch (AGB), its increase in radius can result in the engulfment of its inner planets. The planets spiral into the stellar envelope due to drag and experience a wide range of Mach numbers and values of $\ra/R$. This makes these events an ideal physical application for studies of a rigid gravitating sphere. As such, we focus our parameter space and analysis through the lens of planetary engulfment, though our results are applicable to other problems with similar parameters.

Analytical estimates suggest that engulfed planets of mass $\approx$10 $M_{J}$ (where $M_{J}$ is one Jupiter mass) may eject a portion of the stellar envelope \citep{1984MNRAS.208..763L,1998A&A...335L..85N,2023ApJ...950..128O}. The numerical work of \citet{yarza} confirmed this result, but found that most engulfed planets are instead destroyed through either tidal disruption (for planets which fall within tidal radius of the core) or ram-pressure stripping. Regardless of the mechanism of destruction, the energy that these planets deposit during their inspiral causes a temporary increase in stellar radius and luminosity which may be observable as a transient \citep{2018ApJ...853L...1M,2022MNRAS.511.1330G,engulfmenttransient,2023ApJ...950..128O} at a galactic rate of $\sim$0.1 -- 1 per year \citep{2012MNRAS.425.2778M}.

To determine the ultimate fate of the planet, as well as its energy deposition and eccentricity evolution, it is necessary to study the morphology of the flow near the planet and the resulting forces. \citet{2022MNRAS.513.5465S} integrated the trajectory of an engulfed body to find that dynamical friction can cause the orbital eccentricity to either increase or decrease depending on the equation of state and the local density gradient, and that this could result in binaries exiting the common envelope phase with non-zero eccentricity. Their semi-analytic models used the dynamical friction prescribed by \citet{1999ApJ...513..252O}, which is a classic and well-tested result but leaves open the question of which length scales to choose for the Coulomb logarithm. \citet{yarza} performed similar integrations assuming a circular orbit but determined the drag force using a suite of 3-D hydrodynamical simulations. They interpolated both the dynamical friction and hydrodynamical drag from the results of these simulations but remained agnostic as to their functional forms.

Despite the strides made in recent years, there are still several open questions about gravitating spheres such as the dynamical friction on low-mass objects and the form of the hydrodynamical drag. It is also unclear how the results of T16 generalize to lower masses or into the transonic regime. To this end, we perform a series of hydrodynamical simulations of engulfed bodies over a parameter space tailored to answer these questions.

This paper is organized as follows. We describe the numerical setup and parameters of our wind-tunnel simulations in section \ref{sec:setup}. In section \ref{sec:morphology}, we present our findings on flow morphology, including the shock structure as well as the presence of hydrostatic structures. The gravitational and hydrodynamic forces that result from this morphology are analyzed in section \ref{sec:drag}. We discuss these findings and conclude in section \ref{sec:discussion}.

\section{Numerical Setup} \label{sec:setup}

We perform calculations using the \athena\ code, which is described in detail by \citet{athena++}. \athena\ solves the Euler equations
\be
\ddt{\rho} + \grad\cdot\rho\vel &=& 0, \label{eq:continuity}\\
\ddt{\rho\vel} + \grad\cdot\rho\vel\vel + \grad P &=& -\rho\grad\Phi, \label{eq:momentum}\\
\ddt{\rho e} + \grad\cdot\left(\rho e + P\right)\vel &=& -\rho\vel\cdot\grad\Phi \label{eq:energy}
\ee
for a gas with density $\rho$, fluid velocity $\vel$, gravitational potential $\Phi$, specific energy $e$ and pressure $P$. We model the gas as ideal with specific heat ratio $\gamma=5/3$. A second-order van Leer predictor-corrector scheme is used for time integration with the Courant–Friedrichs–Lewy number set to $\eta_{\rm CFL}=0.3$. We use a Harten–Lax–van Leer contact (HLLC) approximate Riemann solver \citep{1994ShWav...4...25T} and handle spatial reconstruction via a piecewise linear method. \athena\ utilizes a van Leer slope limiter, including the corrections described in \citet{2014JCoPh.270..784M} to keep it total variation diminishing.

We perform ``wind-tunnel'' simulations which model the immediate vicinity of the body. This approach is similar to that of \citet{2015ApJ...803...41M}, \citet{2017ApJ...838...56M} and \citet{2020ApJ...897..130D} which studied common envelope evolution by modeling the engulfed body as a sink particle within a wind tunnel. \citet{yarza} used a similar setup with a reflective boundary rather than a sink. \lp{\sout{We use a spherical-polar geometry with the sphere at the centre of the spherical domain and set the radius of the body to $R=0.1\rsun$, which is typical of Jovian planets.}} \lp{We use a spherical-polar geometry with the body at the centre of the spherical domain. The radius of the body sets a characteristic length scale and is held constant throughout this paper at $R=0.1\rsun$, which is a value typical of Jovian planets. In contrast, the accretion radius $\ra$ is allowed to vary according to the local flow conditions, as we discuss in section \ref{sec:parameters}.} The radius of the simulation domain is $R_{\rm out}=100R$, within which we assume that the gas is initially uniform. We use static mesh refinement, reducing the linear size of the mesh by a factor of 2 within a radius of $20R$ and by a factor of 4 within $r=5R$. Unless otherwise stated, our base grid before refinement is $640\times 40\times 10$ cells in the radial, polar, and azimuthal directions, respectively. This results in $1.6\times 10^{6}$ mesh cells after refinement. In Appendix \ref{sec:convergence}, we perform convergence tests for both the linear resolution and the size of our simulation domain. Because the mass of the body far exceeds that of the gas within the domain (by 1 -- 4 orders of magnitude), we neglect the self-gravity of the gas and treat the sphere as the only source of gravity.

\subsection{Boundary Conditions} \label{sec:BCs}

\begin{figure}
  \includegraphics[height=0.41\textwidth]{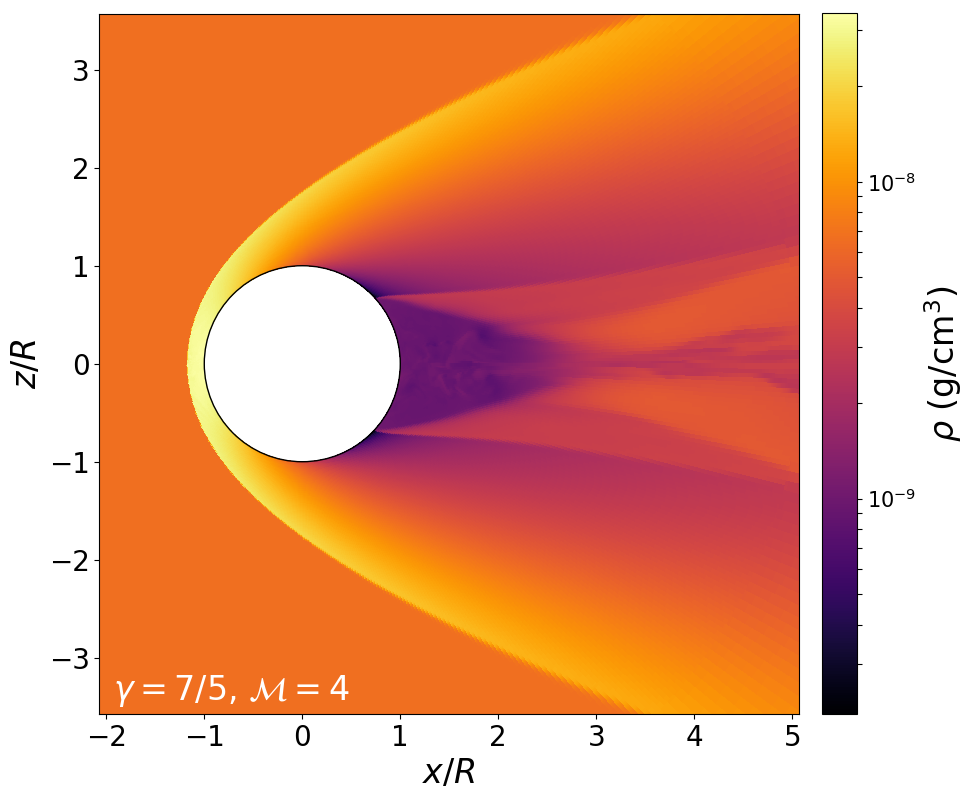}
  \includegraphics[height=0.41\textwidth]{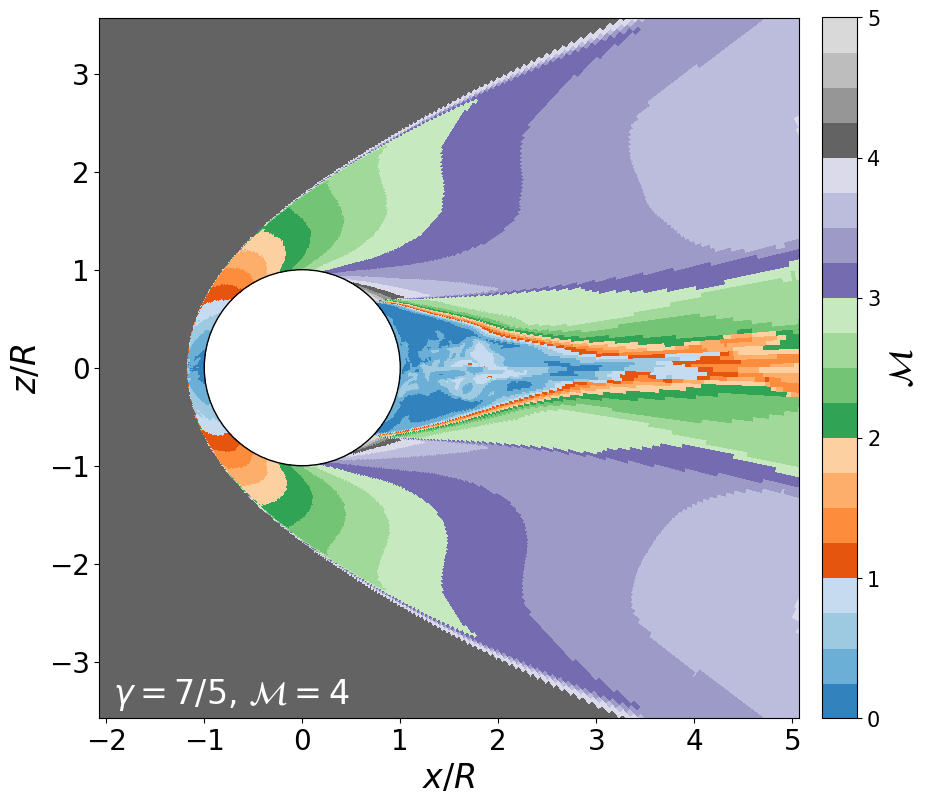}
    \caption{Density and Mach number slices from our $\gamma=7/5$, $\mach=4$ test case in its steady state.
    \label{fig:nograv}}
\end{figure}

\begin{table}
\begin{center}
\caption{Comparison of the shock stand-off distances from our $\gamma=7/5$ test cases ($\rs$) to the fit of \citet{1967JSpRo...4..822B} ($R_{s,B}$) as well as the fractional error.}
\begin{tabular}{c c c c}
\hline
\hfill $\mach$ & $R_{s,B}/R$ & $\rs/R$ & $\Delta$ \\
\hline
\hfill 1.5 & 1.6036 & 1.6125 & $5.58\times10^{-3}$ \\
\hfill 2   & 1.3215 & 1.3539 & $2.46\times10^{-2}$ \\
\hfill 3   & 1.2050 & 1.2178 & $1.06\times10^{-2}$ \\
\hfill 4   & 1.1751 & 1.1837 & $7.32\times10^{-3}$ \\
\hfill 6   & 1.1565 & 1.1565 & $2.86\times10^{-5}$ \\
\hline
\label{tab:lab}
\end{tabular}
\end{center}
\end{table}

The surface of the body is treated as a reflective boundary. We do not impose the no-slip condition at the surface, as we take the Reynolds number to be effectively infinite and the boundary layer to be negligibly small. For the outer boundary, the upstream ($\theta<\pi/2$) and downstream ($\theta>\pi/2$) regions are treated separately:
\begin{enumerate}[(i)]
    \item At the upstream boundary, the fluid is set to some specified inflow conditions $\rho_{\infty}$, $P_{\infty}$ and $\vel_{\infty}$. We assume that the orbital separation is much larger than the domain, so that the curvature of the velocity field is negligible. The incoming streamlines are taken to be parallel so that $\vel_{\infty}=v_{\infty}\hat{x}$, though we investigate the domain size needed for this assumption to be valid in Appendix \ref{sec:convergence}. \\

    \item The downstream boundary is set to ``diode'' conditions: gas is not allowed to enter the domain but may exit with zero-gradient conditions.
\end{enumerate}

We check that this methodology is valid by comparing against the results of laboratory tests. A key quantity in describing a bow shock is the stand-off distance of the nose of the shock from the body $R_{s}$. This is defined as the distance from the centre of the sphere to the bow shock in the upstream direction (along the $\theta=0$ pole). \citet{1967JSpRo...4..822B} complied experimental data from various supersonic wind tunnels to obtain the fit
\be
\frac{R_{s,B}}{R} = 1 + 0.143\exp\left(\frac{3.24}{\mach^{2}}\right).
\ee
We carry out several test simulations with $\mach$ in the range 1.5 -- 6 and set $\gamma=7/5$ to be consistent with the laboratory tests in air. Mach numbers above this range can lead to chemical changes in the air and are not of interest to us. We use a slightly different setup than that described above, using a smaller domain since gravity is absent: the radius of the simulation domain is $R_{\rm out}=50R$ and the base grid is $1800\times 80\times 40$ cells in the radial, polar, and azimuthal directions. We refine this grid by a factor of 2 within $r=10R$ and by 4 within $r=5R$ for a total of $4.27\times10^{7}$ cells. Fig. \ref{fig:nograv} shows density and $\mach$ snapshots from the $\mach=4$ test after it has reached a steady state condition. In Table \ref{tab:lab}, we compare the stand-off distances obtained from our simulation results against the fit of \citet{1967JSpRo...4..822B} through the fractional error
\be
\Delta = \frac{R_{s}-R_{s,B}}{R_{s,B}}.
\ee
We find good agreement between our results and the fit, with errors at or below a few percent.

\subsection{Non-Linearity Parameter} \label{sec:eta}

We perform a suite of simulations in which we vary the mass and Mach number of the body. Here an important quantity is the ``non-linearity parameter''
\be
\eta = \frac{1}{2} \frac{\mach^{2}}{\mach^{2}-1} \frac{R_{A}}{R}, \label{eq:eta}
\ee
as defined by \citet{2009ApJ...703.1278K}, who found that $\eta$ can be used to describe the shock stand-off distance $\rs$ via a broken power law:
\be
\frac{\rs}{R} = \begin{cases} \eta & \eta \gtrsim 2, \\ 2(\eta/2)^{2.8} & 0.7 \lesssim \eta \lesssim 2, \end{cases}
\ee
though they did not perform any runs with $\eta<1$. This result was partially confirmed by T16, who found that 
\be
\frac{\rs}{R} = \begin{cases} \eta & \eta \gtrsim 1, \\ {\rm constant} & \eta \lesssim 1. \end{cases} \label{eq:rsthun}
\ee
T16 also introduced a new formula for dynamical friction when $1<\eta<50$ and demonstrated that it is a good approximation to their numerical results within this range, but that it greatly overestimates the dynamical friction for $\eta<1$. We discuss the dynamical friction further in section \ref{sec:df}.

It is clear that further investigations of the $\eta<1$ regime are necessary. To this end, we perform a suite of simulations with $0\leq\eta\leq1$. We also perform several runs with $\eta>1$ to probe the transition into the high-$\eta$ regime, which is well-understood due to the works mentioned above. We consider only Mach numbers greater than one due to the fact that for subsonic flow, there is no bow shock and no local solution for the hydrostatic halo. Mach numbers greater than 4 are also not considered since they are of little importance to planetary engulfment, as we discuss below.

\subsection{Simulation Realm Applicable to Planetary Engulfment}\label{sec:parameters}

Exoplanet surveys have identified a population of planets orbiting subgiants at separations less than 1 AU which are expected to be engulfed \citep{2011A&A...532A..79S,2011PASP..123..412W,2014ApJ...794....3V}. \citet{2022arXiv221015848L} carried out global simulations of a hot Jupiter engulfed by an early red giant, which is a numerically tractable problem as the inspiral proceeds on the dynamical timescale of the star. However, many planets are expected to be engulfed at a later evolutionary stage where the envelope has expanded to hundreds of solar radii, requiring many hundreds of orbits to inspiral. This motivates our assumption that the gas within the simulation domain is uniform, as for near-sonic $\mach$ the local density scale height is comparable to the orbital separation.

To set the stage, we choose conditions typical of the outer envelope of an AGB star: $\gamma=5/3$, $\rho_{\infty}=6.76\times 10^{-9}$ g/cm$^{3}$ and $P_{\infty}=9109$ dyn/cm$^{2}$. The velocity $v_{\infty}$ is then determined from the specified Mach number $\mach_{\infty}$. The choices of $\rho_{\infty}$ and $P_{\infty}$ do not affect the flow morphology except for their role in determining the sound speed $c_{s,\infty}=\sqrt{\gamma P_{\infty}/\rho_{\infty}}$.

The Mach number of an engulfed planet in a convective envelope is generally expected to be supersonic, though subsonic velocities are possible. Consider the specific energy of (static) stellar material
\be
e = -\frac{G m_{\rm enc}}{r} + \epsilon,
\ee
where $m_{\rm enc}$ is the stellar mass enclosed within radius $r$ and the specific internal energy $\epsilon$ can be expressed as
\be
\epsilon = \frac{1}{\gamma-1} \frac{P}{\rho} = \frac{c_{s}^{2}}{\gamma(\gamma-1)}.
\ee
For a circular orbit appropriate to a slow inspiral, $G m_{\rm enc}/r = v^{2}$. Then the specific energy of the stellar material is
\be
e = -v^{2} + \frac{c_{s}^{2}}{\gamma(\gamma-1)}.
\ee
If the stellar material is locally bound ($e<0$), we arrive at the condition
\be
\mach^{2} > \frac{1}{\gamma(\gamma-1)} = \frac{9}{10}. \label{eq:machcondition}
\ee
\citet{2023ApJ...950..128O} find that in practice the Mach number always exceeds unity for a range of AGB stellar models generated in the stellar evolution code \mesa\ \citep{2011ApJS..192....3P,2013ApJS..208....4P,2015ApJS..220...15P,2018ApJS..234...34P,2019ApJS..243...10P,2023ApJS..265...15J}. Notably, they also find that $\eta$ falls in the range $0.05\lesssim\eta\lesssim2$ for a Jupiter-mass planet. However, \citet{2022MNRAS.513.5465S} find subsonic behavior over a portion of the inner envelope for a range of $\gamma$ values including $\gamma=5/3$, using solutions of the Lane-Emden equation to obtain the stellar profile. This discrepancy in $\mach$ could be attributable to differences in stellar structure between these two works, though they also use different drag prescriptions. Regardless, it is clear that the Mach number is low yet supersonic over a large portion of the envelope.

Based on the considerations presented in this section, we choose a parameter space of $1.25\leq\mach\leq4$ and $0\leq\eta\leq2$, as well as several runs with higher $\eta$ and $\mach=4$. It is necessary to carry out several runs with varying $\mach$ for each value of $\eta$, as there is no guarantee that the flow morphology and drag forces depend solely on $\eta$. The alert reader will note from (\ref{eq:eta}) that for fixed $\eta$, an increase in $\mach$ corresponds to an increase in $\ra$. A summary of the initial conditions and derived quantities for our calculations can be found in Table \ref{tab:setups}.

\begin{table*}
\begin{center}
\caption{Input parameters and derived quantities for our calculations performed in this work. From left to right, the quantities are: non-linearity parameter $\eta$, inflow Mach number $\mach_{\infty}$, accretion radius $\ra$, mass of the gravitating body $m$ in units of Jupiter mass $M_{J}$, specific heat ratio $\gamma$, radius of simulation domain $R_{\rm out}$, shock stand-off distance $\rs$, boundary of the separation bubble $\theta_{b}$, pseudoentropy in the separation bubble or halo $\sigma_{b}$, pressure drag coefficient $C_{p}$, gravitational drag coefficient $C_{g}$ measured at $r=20R$ and effective linear size of the perturber $\smin$. The sound speed $c_{s,\infty}=\sqrt{\gamma P_{\infty}/\rho_{\infty}}$ depends only on $\gamma$, as $\rho_{\infty}$ and $P_{\infty}$ are constant.}
\begin{tabular}{c c c c c c c c c c c c c c}
\hline
\hfill $\eta$ & $\mach_{\infty}$ & $R_{A}/R$ & $m/M_{J}$ & $\gamma$ & $R_{\rm out}/R$ & $\rs/R$ & $\theta_{b}$ (deg) & $\sigma_{b}/\sigma_{\infty}$ & $C_{p}$ & $C_{g}|_{20R}$ & $s_{\rm min}$ \\
\hline
\hfill 0   & 1.5  & 0     & 0      & 7/5 & 50   & 1.612 & 129 & 1.413 &  1.044 & 0 & -- \\
\hfill 0   & 2    & 0     & 0      & 7/5 & 50   & 1.354 & 131 & 2.588 &  1.018 & 0 & -- \\
\hfill 0   & 3    & 0     & 0      & 7/5 & 50   & 1.218 & 135 & 4.961 &  0.955 & 0 & -- \\
\hfill 0   & 4    & 0     & 0      & 7/5 & 50   & 1.184 & 138 & 16.03 &  0.919 & 0 & -- \\
\hfill 0   & 6    & 0     & 0      & 7/5 & 50   & 1.156 & 138 & 38.83 &  0.891 & 0 & -- \\
\hline
\hfill 0   & 1.25 & 0     & 0      & 5/3 & 100  & 2.199 & 125 & 2.419 &  0.879 & 0 & -- \\
\hfill 0   & 1.5  & 0     & 0      & 5/3 & 100  & 1.735 & 129 & 3.366 &  0.923 & 0 & -- \\
\hfill 0   & 1.75 & 0     & 0      & 5/3 & 100  & 1.542 & 130 & 2.438 &  0.931 & 0 & -- \\
\hfill 0   & 2    & 0     & 0      & 5/3 & 100  & 1.464 & 131 & 5.860 &  0.920 & 0 & -- \\
\hfill 0   & 2.5  & 0     & 0      & 5/3 & 100  & 1.348 & 133 & 9.572 &  0.906 & 0 & -- \\
\hfill 0   & 3    & 0     & 0      & 5/3 & 100  & 1.309 & 133 & 14.54 &  0.884 & 0 & -- \\
\hfill 0   & 3.5  & 0     & 0      & 5/3 & 100  & 1.271 & 134 & 20.67 &  0.873 & 0 & -- \\
\hfill 0   & 4    & 0     & 0      & 5/3 & 100  & 1.271 & 137 & 24.57 &  0.861 & 0 & -- \\
\hline
\hfill 0.4 & 1.25 & 0.288 & 0.0279 & 5/3 & 100  & 2.044 & 127 & 1.990 &  1.004 & -0.724 & 41.3 \\
\hfill 0.4 & 1.5  & 0.444 & 0.0621 & 5/3 & 100  & 1.619 & 129 & 2.218 &  1.142 & 0.676 & 10.2 \\
\hfill 0.4 & 1.75 & 0.539 & 0.102  & 5/3 & 100  & 1.464 & 129 & 2.479 &  1.217 & 1.149 & 6.34 \\
\hfill 0.4 & 2    & 0.600 & 0.149  & 5/3 & 100  & 1.387 & 129 & 2.862 &  1.265 & 1.382 & 5.02 \\
\hfill 0.4 & 2.5  & 0.672 & 0.261  & 5/3 & 100  & 1.348 & 129 & 3.745 &  1.313 & 1.615 & 3.98 \\
\hfill 0.4 & 3    & 0.711 & 0.397  & 5/3 & 100  & 1.310 & 129 & 5.005 &  1.339 & 1.725 & 3.56 \\
\hfill 0.4 & 3.5  & 0.735 & 0.559  & 5/3 & 100  & 1.271 & 130 & 6.485 &  1.366 & 1.782 & 3.37 \\
\hfill 0.4 & 4    & 0.750 & 0.745  & 5/3 & 100  & 1.271 & 130 & 8.260 &  1.379 & 1.816 & 3.25 \\
\hline
\hfill 0.6 & 1.25 & 0.432 & 0.0419 & 5/3 & 100  & 1.967 & 127 & 1.785 &  1.038 & 0.342 & 14.2 \\
\hfill 0.6 & 1.5  & 0.667 & 0.0931 & 5/3 & 100  & 1.580 & 128 & 1.915 &  1.159 & 1.345 & 5.21 \\
\hfill 0.6 & 1.75 & 0.808 & 0.154  & 5/3 & 100  & 1.464 & 127 & 2.182 &  1.187 & 1.684 & 3.71 \\
\hfill 0.6 & 2    & 0.900 & 0.223  & 5/3 & 100  & 1.387 & 125 & 2.476 &  1.211 & 1.861 & 3.11 \\
\hfill 0.6 & 2.5  & 1.01  & 0.391  & 5/3 & 100  & 1.310 & 124 & 3.254 &  1.201 & 2.051 & 2.57 \\
\hfill 0.6 & 3    & 1.07  & 0.596  & 5/3 & 100  & 1.310 & 123 & 4.312 &  1.170 & 2.145 & 2.34 \\
\hfill 0.6 & 3.5  & 1.10  & 0.838  & 5/3 & 100  & 1.271 & 123 & 5.549 &  1.153 & 2.200 & 2.22 \\
\hfill 0.6 & 4    & 1.12  & 1.12   & 5/3 & 100  & 1.271 & 122 & 7.050 &  1.130 & 2.231 & 2.15 \\
\hline
\hfill 1   & 1.25 & 0.720 & 0.0698 & 5/3 & 100  & 1.851 & 124  & 1.549 &  0.957 & 1.276 & 5.58 \\
\hfill 1   & 1.5  & 1.11  & 0.155  & 5/3 & 100  & 1.542 & 120  & 1.604 &  0.765 & 1.969 & 2.79 \\
\hfill 1   & 1.75 & 1.35  & 0.256  & 5/3 & 100  & 1.426 & 116  & 1.806 &  0.537 & 2.212 & 2.19 \\
\hfill 1   & 2    & 1.50  & 0.372  & 5/3 & 100  & 1.387 & 114  & 2.105 &  0.284 & 2.354 & 1.90 \\
\hfill 1   & 2.5  & 1.68  & 0.652  & 5/3 & 100  & 1.310 & 107  & 2.814 & -0.205 & 2.521 & 1.61 \\
\hfill 1   & 3    & 1.78  & 0.993  & 5/3 & 100  & 1.310 & 103  & 3.751 & -0.584 & 2.614 & 1.46 \\
\hfill 1   & 3.5  & 1.84  & 1.40   & 5/3 & 100  & 1.271 & 101  & 4.800 & -0.840 & 2.674 & 1.38 \\
\hfill 1   & 4    & 1.88  & 1.86   & 5/3 & 100  & 1.271 & 98.4 & 6.053 & -0.999 & 2.706 & 1.34 \\
\hline
\hfill 1   & 4   & 1.88  & 1.86    & 5/3 & 10   & 1.271 & 86.1 & 5.695 & -1.535 & 2.006 & 2.69 \\
\hfill 1   & 4   & 1.88  & 1.86    & 5/3 & 20   & 1.268 & 93.9 & 5.975 & -1.318 & 2.638 & 1.43 \\
\hfill 1   & 4   & 1.88  & 1.86    & 5/3 & 50   & 1.268 & 97.3 & 5.807 & -1.174 & 2.748 & 1.28 \\
\hfill 1   & 4   & 1.88  & 1.86    & 5/3 & 150  & 1.272 & 98.4 & 6.297 & -0.959 & 2.694 & 1.35 \\
\hfill 1   & 4   & 1.88  & 1.86    & 5/3 & 200  & 1.272 & 98.4 & 6.081 & -0.967 & 2.690 & 1.36 \\
\hline
\hfill 2   & 1.25 & 1.44  & 0.140  & 5/3 & 100  & 1.851 & -- & 1.096 & -0.179 & 1.731 & 3.54 \\
\hfill 2   & 1.5  & 2.22  & 0.310  & 5/3 & 100  & 1.658 & -- & 1.344 & -0.530 & 1.983 & 2.75 \\
\hfill 2   & 1.75 & 2.69  & 0.512  & 5/3 & 100  & 1.658 & -- & 1.608 & -0.404 & 2.061 & 2.55 \\
\hfill 2   & 2    & 3.00  & 0.745  & 5/3 & 100  & 1.735 & -- & 1.944 & -0.283 & 2.074 & 2.51 \\
\hfill 2   & 2.5  & 3.36  & 1.30   & 5/3 & 100  & 1.928 & -- & 2.677 & -0.247 & 2.090 & 2.47 \\
\hfill 2   & 3    & 3.56  & 1.99   & 5/3 & 100  & 2.045 & -- & 3.489 & -0.158 & 2.078 & 2.50 \\
\hfill 2   & 3.5  & 3.67  & 2.79   & 5/3 & 100  & 2.161 & -- & 4.429 &  0.291 & 2.061 & 2.55 \\
\hfill 2   & 4    & 3.75  & 3.72   & 5/3 & 100  & 2.199 & -- & 5.708 &  0.194 & 2.032 & 2.62 \\
\hline
\hfill 4   & 4    & 7.50  & 7.45   & 5/3 & 400  & 4.104 & -- & 4.137 & -0.079 & 2.761 & 5.06 \\
\hline
\hfill 7   & 4    & 13.1  & 13.0   & 5/3 & 700  & 7.111 & -- & 5.800 &  0.158 & 2.768 & 8.79 \\
\hline
\hfill 10  & 4    & 18.8  & 18.6   & 5/3 & 1000 & 10.03 & -- & 5.833 & -0.076 & 1.846 & 12.6 \\
\hline
\label{tab:setups}
\end{tabular}
\end{center}
\end{table*}

\section{Flow Morphology} \label{sec:morphology}

\begin{figure*}
  \includegraphics[width=0.95\textwidth]{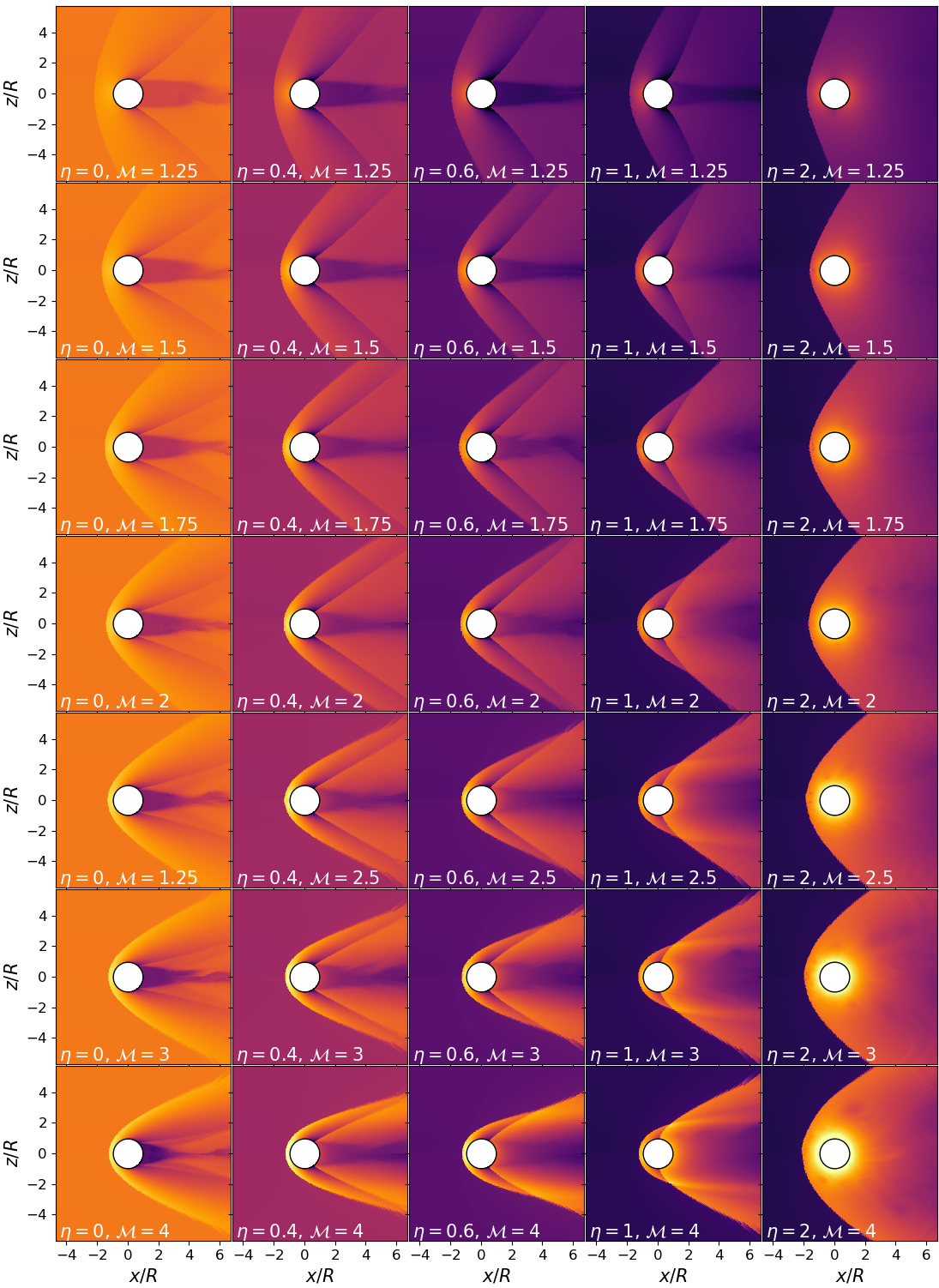}
    \caption{Density snapshots from our $\gamma=5/3$ runs with $\eta\leq2$, omitting $\mach=3.5$ for space. The colormap bounds for each column are unique for the purpose of illustrating the morphology.
    \label{fig:panels}}
\end{figure*}

\begin{figure*}
  \includegraphics[width=1.0\textwidth]{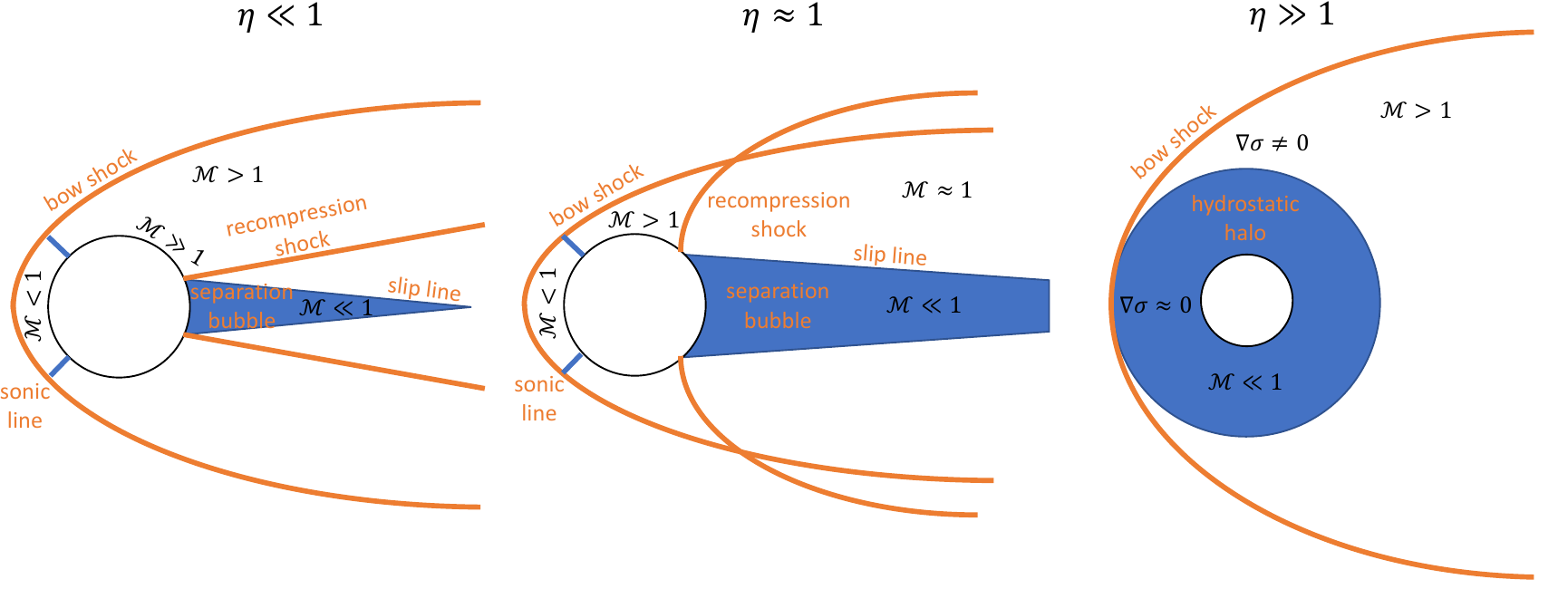}
    \caption{Illustration of the qualitative features of our results for low, intermediate and high $\eta$. For low $\eta$ \textit{(left)}, the morphology resembles the non-gravitating case, with a small separation bubble on the trailing edge. As $\eta$ is increased \textit{(centre)}, the separation bubble increases in size and in its extent along the surface of the sphere, driving the recompression shock outward to intersect with the bow shock. Further increase in $\eta$ \textit{(right)} causes the separation bubble to envelop the body and transition into an isentropic, hydrostatic halo.
    \label{fig:doubleshock}}
\end{figure*}

Each simulation is run until it reaches steady state, as determined by the time variation of its derived quantities. This takes approximately one fluid crossing time of the domain $t_{cr}=2R_{\rm out}/v_{\infty}$. Typically a vortex ring briefly forms behind the sphere before being shed downstream. The exception is our non-gravitating $\mach=1.5$ test case with $\gamma=7/5$, which retains its vortex ring. Density snapshots from our $\gamma=5/3$ runs are shown in Fig. \ref{fig:panels}, with $\eta$ increasing to the right and $\mach$ increasing down each column. The colormap bounds are unique to each column for the purpose of illustrating flow morphology. These snapshots show the $\phi=0$ plane, though we do not find significant azimuthal dependence in any of our 3-D results. The qualitative features of these results are illustrated in Fig. \ref{fig:doubleshock} as a primer for the following discussion of these features.

\subsection{Detached Shock Stand-Off} \label{sec:standoff}

\begin{figure}
  \includegraphics[width=0.5\textwidth]{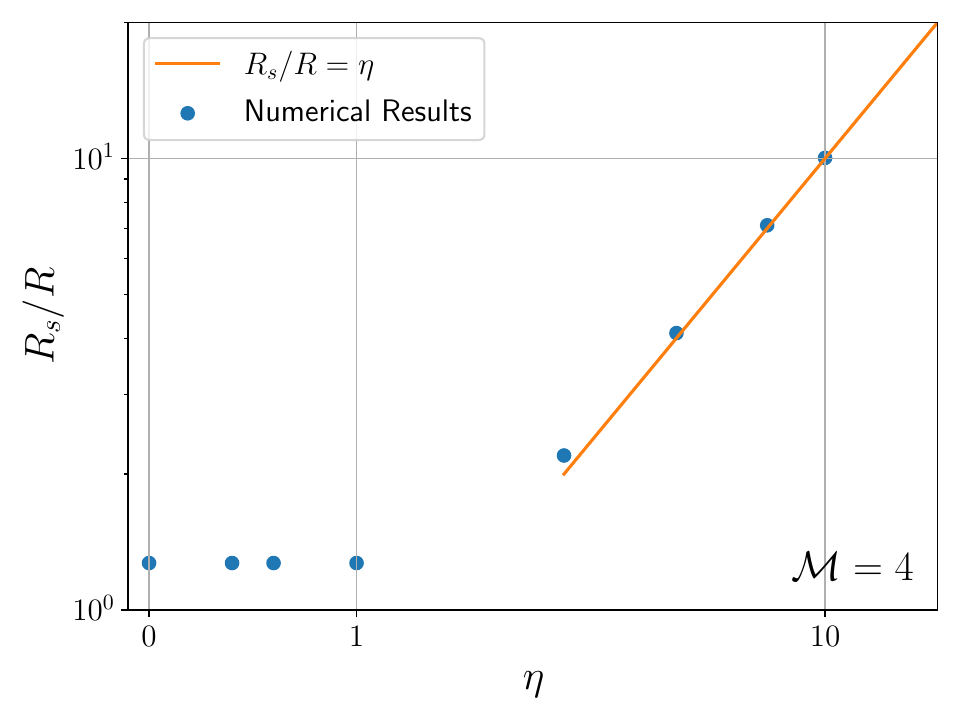}
    \caption{Variation of the shock stand-off distance $\rs$ with the non-linearity parameter $\eta$ for $\mach=4$. Our results (blue dots) are consistent with the result of T16 that $\rs=\eta R$ for $\eta>1$ (orange line) and constant for $\eta\leq1$.
    \label{fig:etarso}}
\end{figure}

\begin{figure}
  \includegraphics[width=0.5\textwidth]{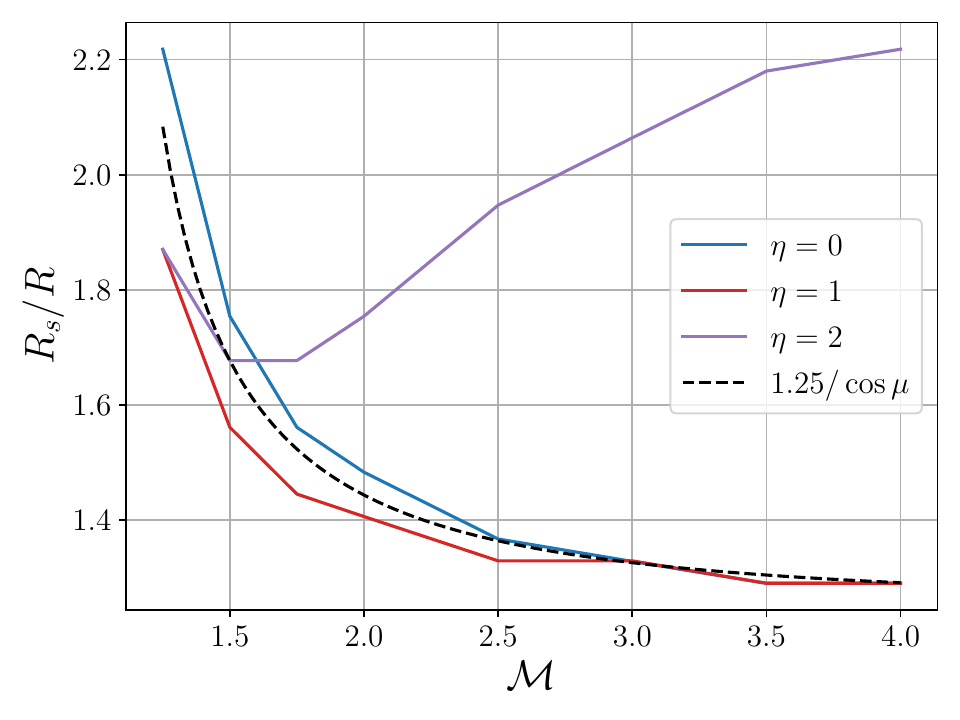}
    \caption{Shock stand-off distance $\rs$ as a function of $\mach$ for several values of $\eta$. Significant deviations from the high-$\mach$ value are seen for $\mach$ approaching unity, which scales with the cosine of the Mach angle.
    \label{fig:rsovsmach}}
\end{figure}


In Fig. \ref{fig:etarso} we show the shock stand-off distance as a function of $\eta$ for our $\mach=4$ runs. We find that $R_{s}/R=\eta$ above $\eta=1$ and is constant below, in agreement with T16. However, when we decrease the Mach number toward unity we find that $R_{s}$ depends strongly on $\mach$, as shown in Fig. \ref{fig:rsovsmach}. Specifically, it is proportional to the cosine of the Mach angle $\cos\mu=\sqrt{\mach^{2}-1}/\mach$, though its scaling with $\eta$ is ambiguous. On the other hand, $\eta=2$ exhibits altogether different behavior, generally increasing $\rs$ with $\mach$. Because an increase in $\mach$ corresponds to an increase in $\ra$ at fixed $\eta$, it is not guaranteed that increasing $\mach$ will decrease $\rs$ in contrast to the non-gravitating case.

The variation of $\rs$ for a single value of $\eta$ calls into question the usefulness of the $\eta$ parameter as a metric, as opposed to the accretion radius. However, we find that cases with $\eta\leq1$ exhibit a distinctly different flow morphology than those with $\eta>1$ despite overlapping in $\ra$, as we now discuss.

\subsection{Shock Structure} \label{sec:structure}

\begin{figure}
  \includegraphics[width=0.5\textwidth]{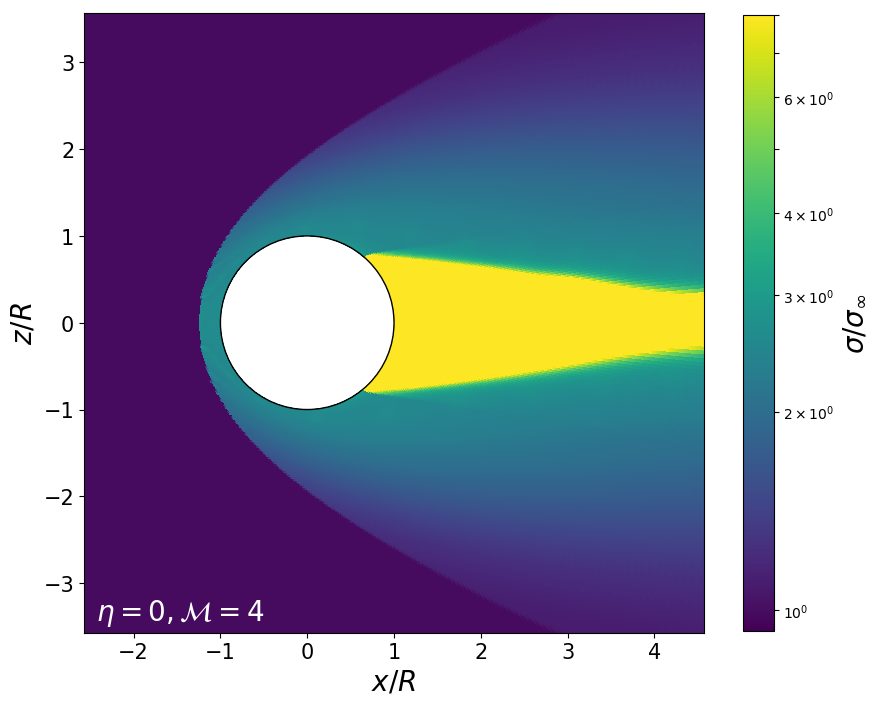}
  \includegraphics[width=0.5\textwidth]{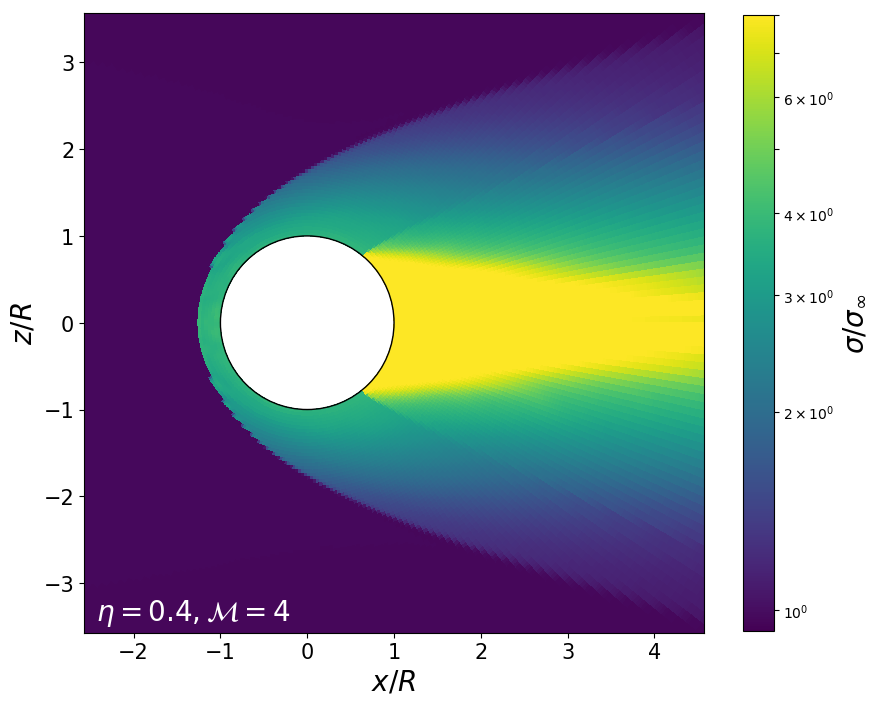}
  \includegraphics[width=0.5\textwidth]{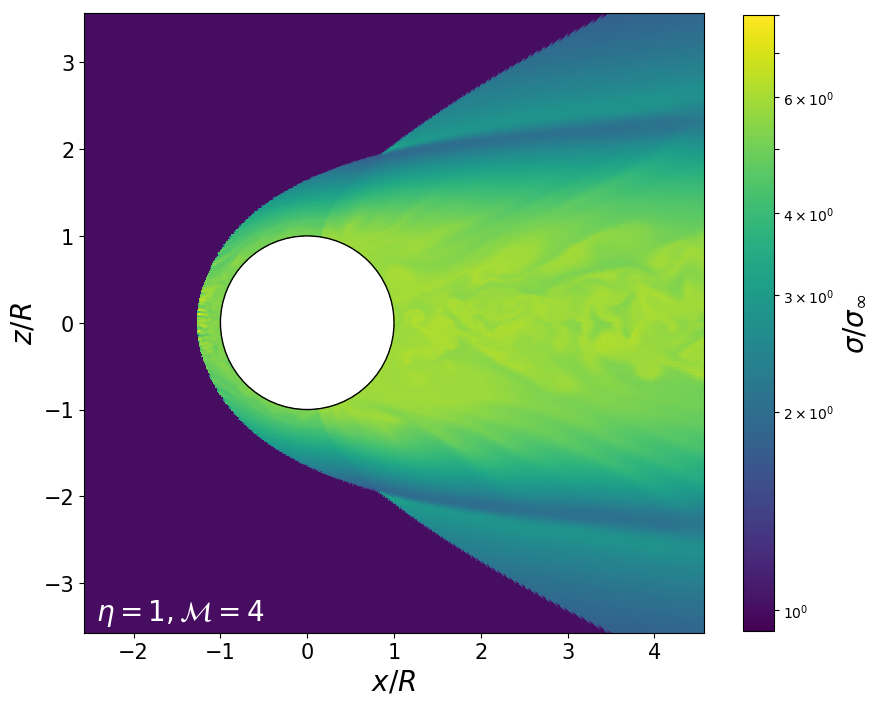}
    \caption{Entropy slices for $\eta=0$ \textit{(top)}, $\eta=0.4$ \textit{(middle)} and $\eta=1$ \textit{(bottom)} with $\mach=4$, normalized to the entropy of the inflowing gas. The entropy of the separation bubble is much higher than the surrounding gas. All plots have identical colormap limits for the purpose of comparison.
    \label{fig:entropy}}
\end{figure}

\begin{figure*}
\begin{tabular}{ll}
  \includegraphics[height=0.39\textwidth]{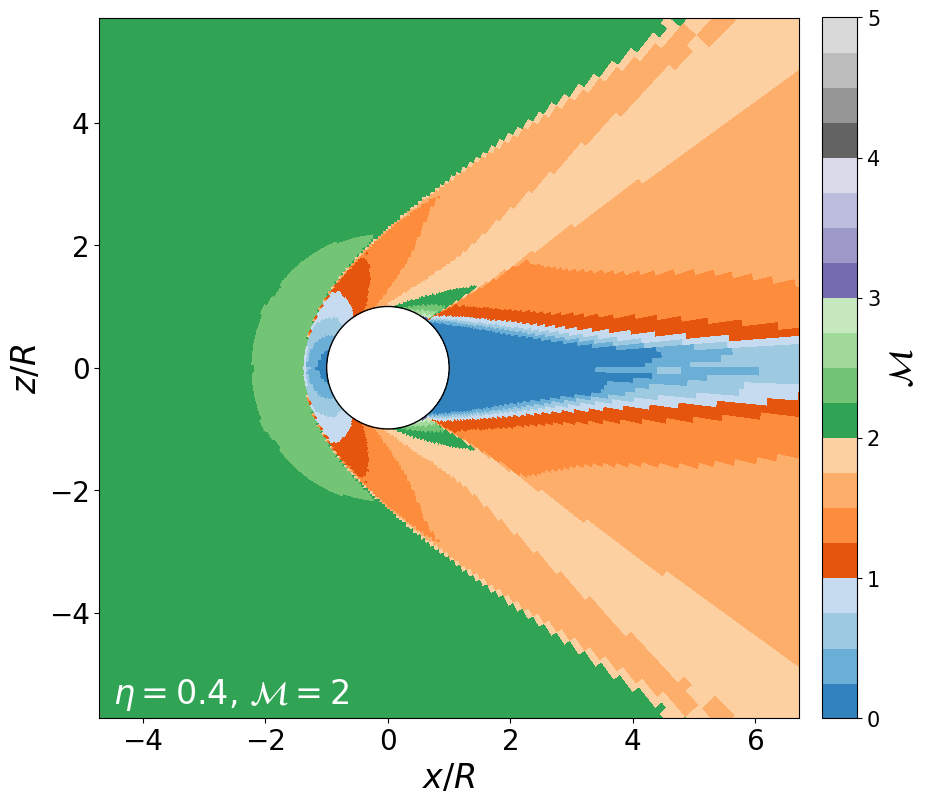} & \includegraphics[height=0.39\textwidth]{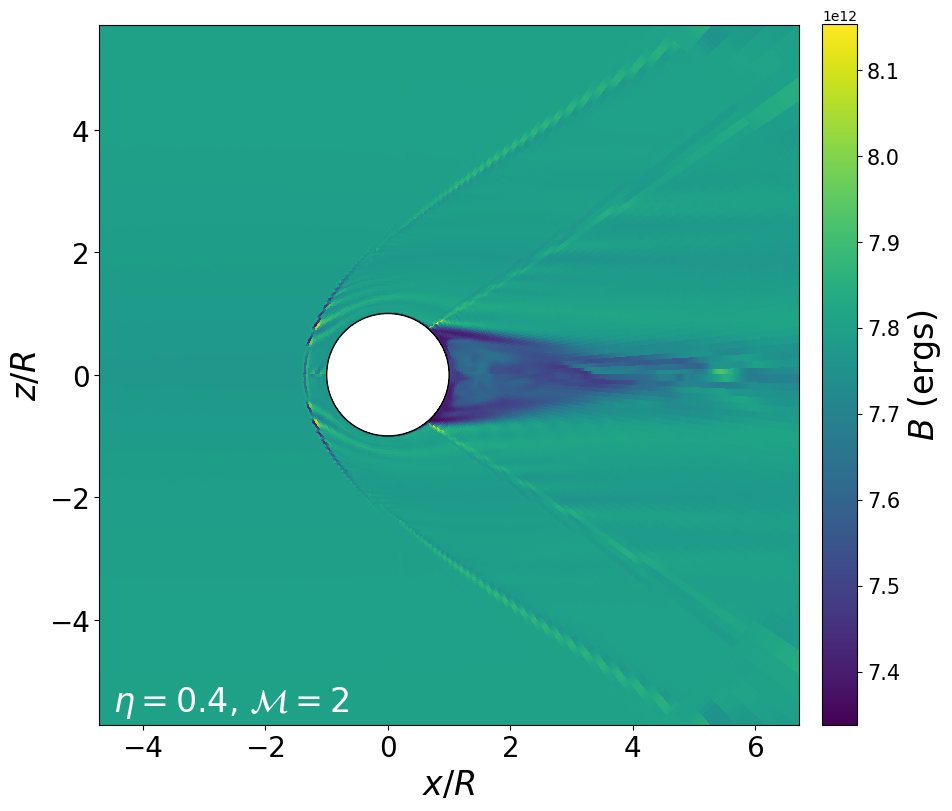} \\
 \includegraphics[height=0.39\textwidth]{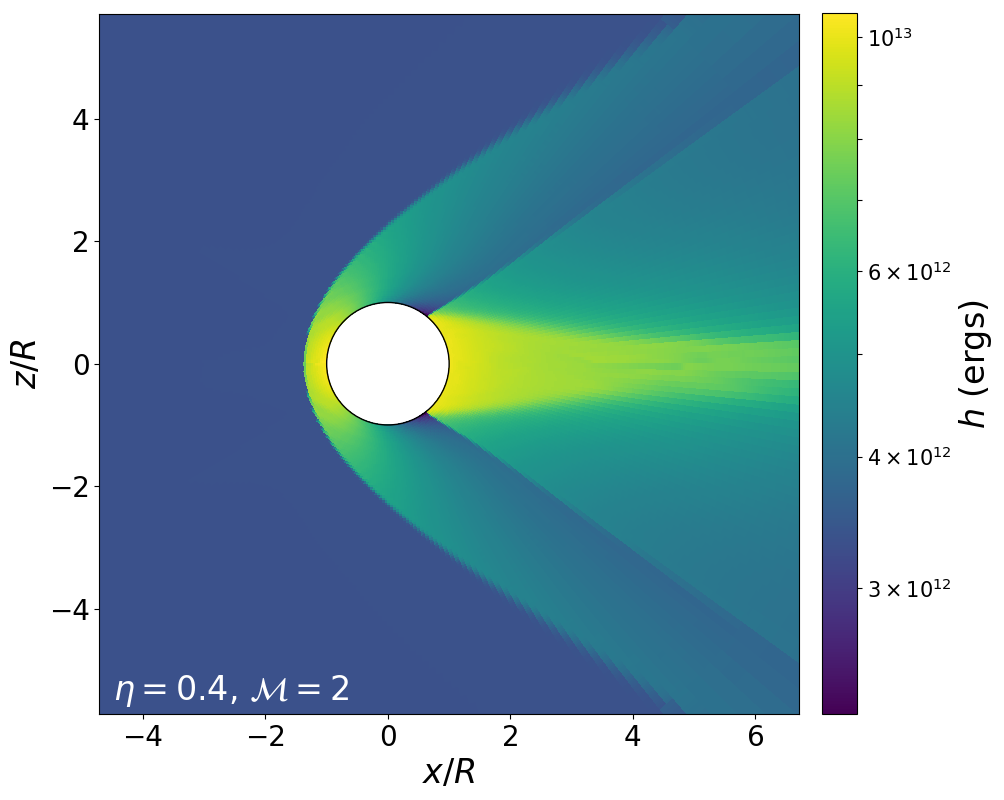} & \includegraphics[height=0.39\textwidth]{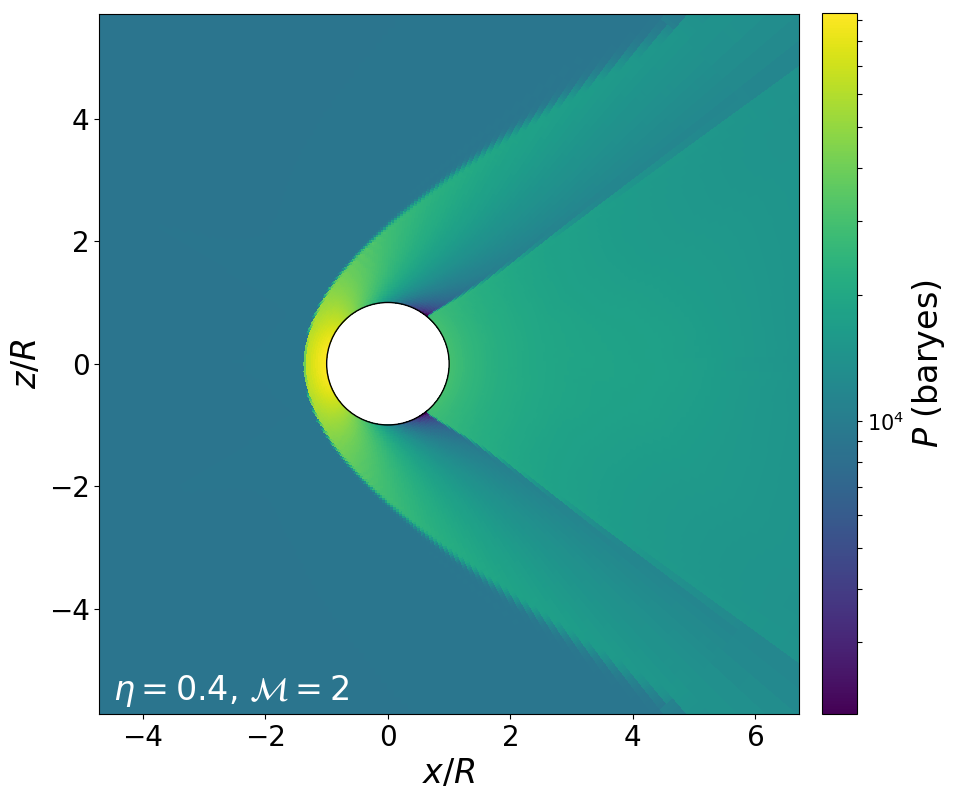} \\
\end{tabular}
\caption{Slices of various quantities for $\eta=0.4$, $\mach=2$. The Mach number \textit{(top left)} contrasts regions of the flow which are subsonic (blue), supersonic but less than the inflow Mach number (orange) and exceeding the inflow Mach number (green). Several features of the morphology are visible such as the sonic line, supersonic expansion just upstream of the recompression shock and near-zero velocities within the separation bubble. The Bernoulli constant \textit{(top right)} is uniform over the domain except for the separation bubble, indicating that it is isolated from the rest of the flow. The enthalpy \textit{(bottom left)} is higher within the bubble than outside the bubble despite the pressure \textit{(bottom right)} being continuous across the contact discontinuity.
\label{fig:eta04mach2}}
\end{figure*}

\begin{figure}
  \includegraphics[height=0.39\textwidth]{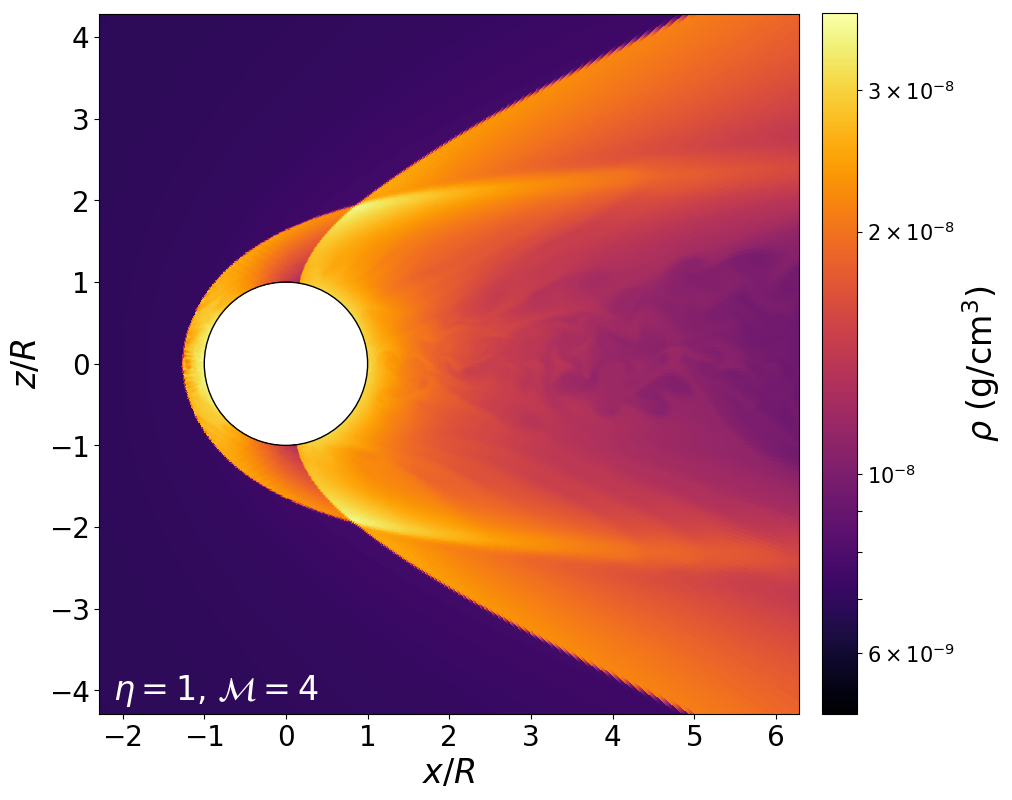}
  \includegraphics[height=0.39\textwidth]{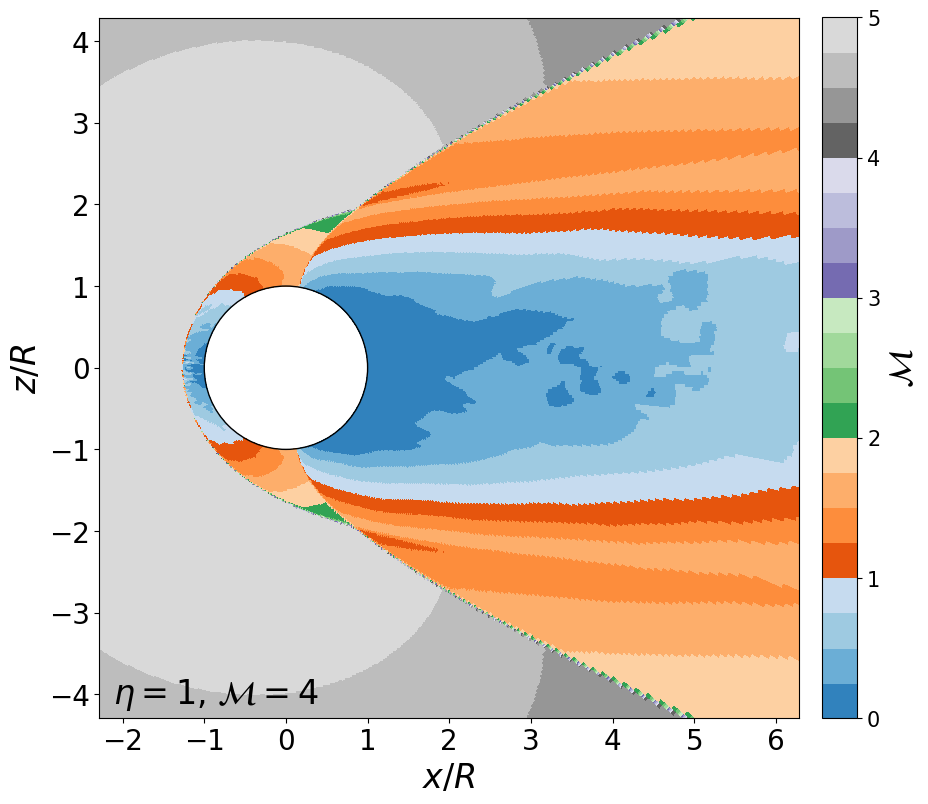}
    \caption{Snapshots of density \textit{(top)} and Mach number \textit{(bottom)} for $\eta=1$, $\mach=4$ showing the shock interaction.
    \label{fig:doubleshockslice}}
\end{figure}

\begin{figure}
  \includegraphics[width=0.5\textwidth]{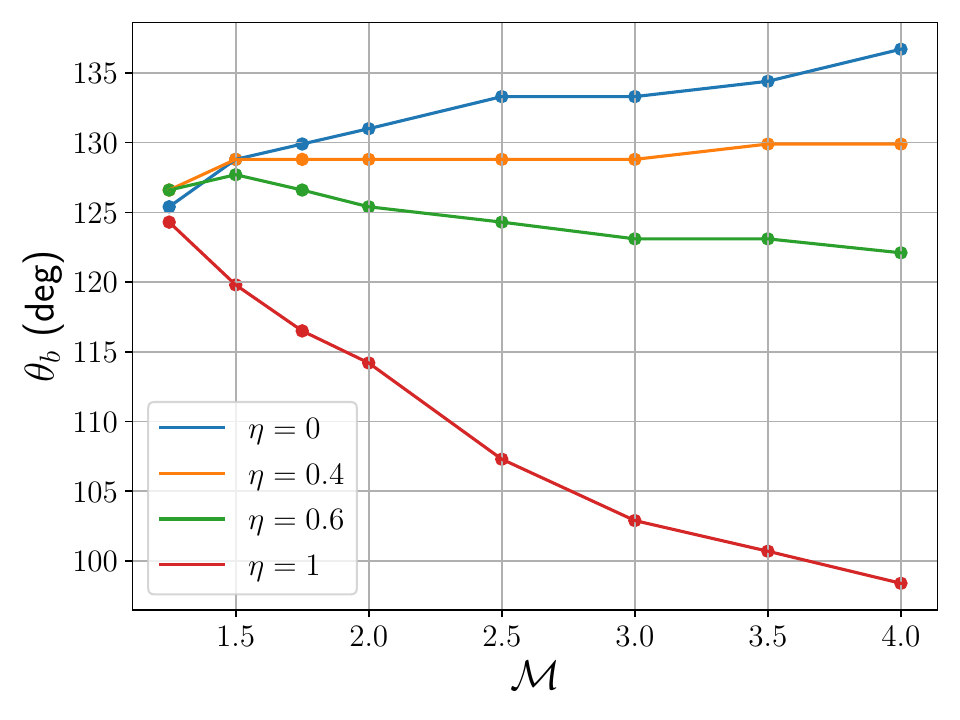}
    \caption{The boundary of the separation bubble along the surface of the sphere ($\theta=0$ is along the upstream direction). For $\eta=0$ we reproduce the expected result that the separation bubble shrinks with increasing $\mach$. As $\eta$ is increased, this trend reverses.
    \label{fig:thetabubble}}
\end{figure}

It was shown above that the stand-off distance can be bifurcated into two distinct regions separated by $\eta=1$. We find that the reason for this is that the flow morphology is also distinctly different in these regions.

At low $\eta$, the bow shock is near the surface of the body, with a shape that is largely determined by the upstream Mach number. Trailing the sphere is a wake comprised of a low-density bubble sheathed in a second shock, visible in many of the panels of Fig. \ref{fig:panels}. The bubble is quite high in entropy and is fairly isentropic, as shown in Fig. \ref{fig:entropy}. There is a large contrast in velocity between the bubble and the rest of the wake, separated by a slip line. This phenomenon also occurs in the absence of gravity and is well-understood: fluid which crosses the bow shock at near its strongest point at its nose is slowed to subsonic speeds, though this region of subsonic flow in front of the body is fairly small. \lp{Indeed, our runs with $\eta=0$ (in which gravity is absent) strongly resemble the non-gravitating test cases of section \ref{sec:BCs}, differing only in the value of $\gamma$.} Streamlines crossing the outer oblique regions of the bow shock remain supersonic, and it is these streamlines which crest the top of the sphere ($\theta=\pi/2$) and undergo supersonic expansion until they separate from its surface. Increasing the Mach number delays boundary layer separation and thus decreases the size of the separation bubble \citep{doi:10.2514/1.28943}. The turning of the streamlines as they encounter the separation bubble necessitates a recompression shock which slows the fluid to near-sonic (or possibly subsonic) speed.

As a case study, consider our run with $\eta=0.4$ and $\mach=2$, which we show in Fig. \ref{fig:eta04mach2}. The top left panel shows the Mach number, which exhibits the features discussed above. To quantify the isolation of the bubble from the rest of the flow, we use the Bernoulli constant
\be
B \equiv h + \frac{v^{2}}{2} + \Phi,
\ee
where
\be
h = \frac{\gamma}{\gamma-1} \frac{P}{\rho},
\ee
is the enthalpy. $B$ is conserved along each streamline in steady flow as a consequence of Crocco's theorem \citep{2000hsf..book.....C}. Furthermore, since the Bernoulli constant is uniform over the inflow boundary, it is also uniform for all flow originating from that boundary. We see in the top right panel that $B$ is constant except for the separation bubble, indicating that the bubble is isolated from the rest of the flow. It is also significantly hotter than the surrounding gas (bottom left panel) while the pressure is continuous across the slip line (bottom right panel). This pressure can exert significant force on the body, which is discussed further in section \ref{sec:pressure}.

As $\eta$ is increased from zero to unity, the gravitational influence of the body causes the wake to increase in size dramatically. This widening of the wake can lead to an intersection between the two shocks, which is particularly pronounced for $\eta=1$ at high $\mach$ (Figs. \ref{fig:entropy} and \ref{fig:doubleshockslice}). In Fig. \ref{fig:thetabubble} we show the boundary of the separation bubble along the surface of the sphere $\theta_{b}$, recalling that $\theta=0$ corresponds to the upstream direction. We see that the non-gravitating result -- that increasing $\mach$ reduces the size of the bubble -- reverses as $\eta$ is increased. In fact, for $\eta=1$, $\mach=4$ the bubble nearly crests the top of the sphere ($\theta=\pi/2$). If $\eta$ is further increased beyond unity, this double-shock structure disappears entirely (shown in the rightmost panels of Fig. \ref{fig:panels}) as the bubble is replaced by a hydrostatic halo, as we show below. We see that the transition is characterized by $\eta$ rather than $\ra$, as $\eta=1$ and $\eta=2$ exhibit different morphologies despite overlapping in their coverage of $\ra$.

\subsection{Hydrostatic Structures} \label{sec:halo}

\begin{figure}
  \includegraphics[width=0.5\textwidth]{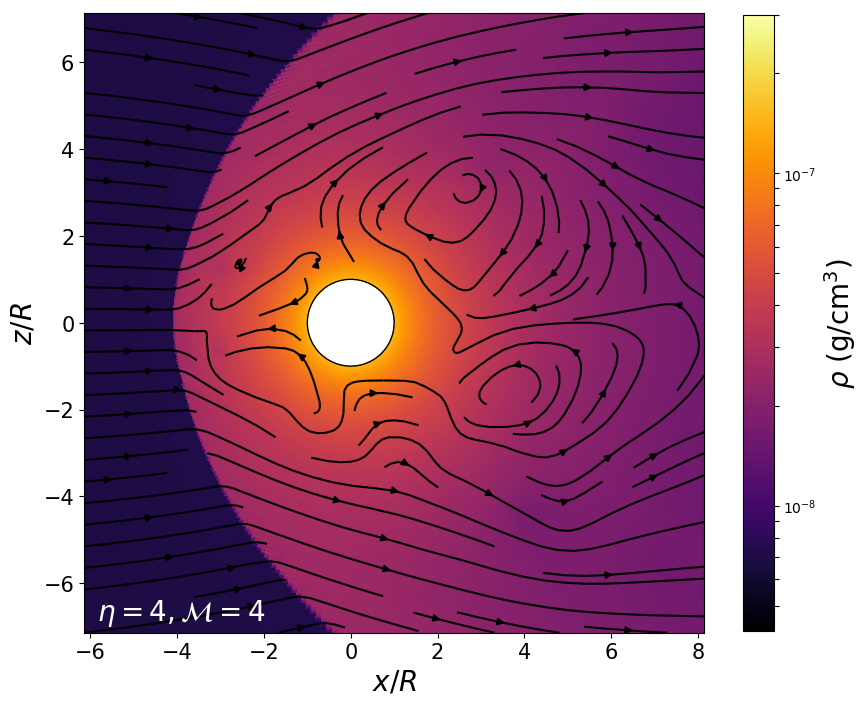}
  \includegraphics[width=0.5\textwidth]{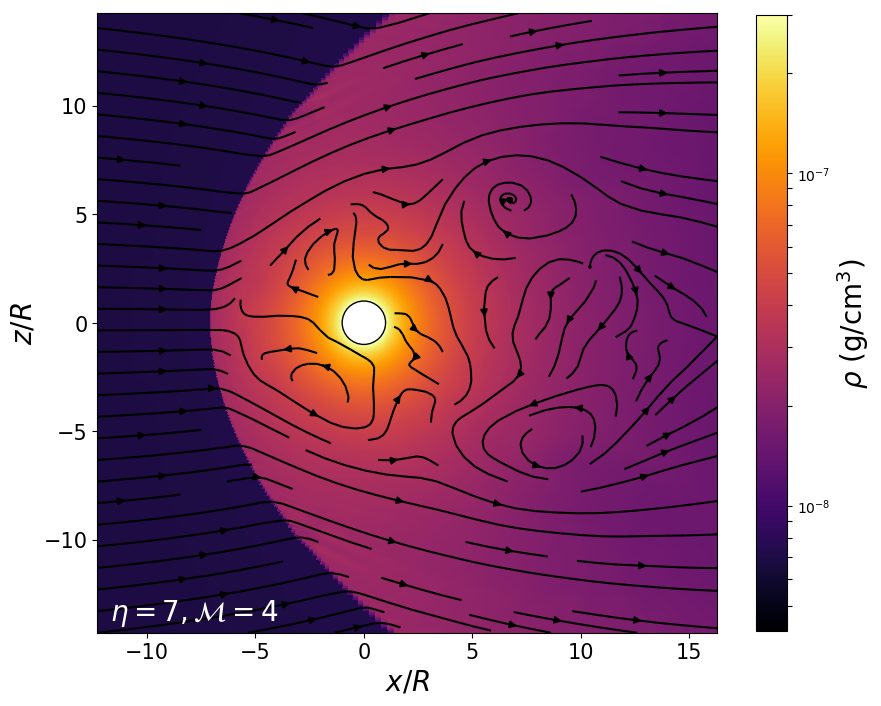}
  \includegraphics[width=0.5\textwidth]{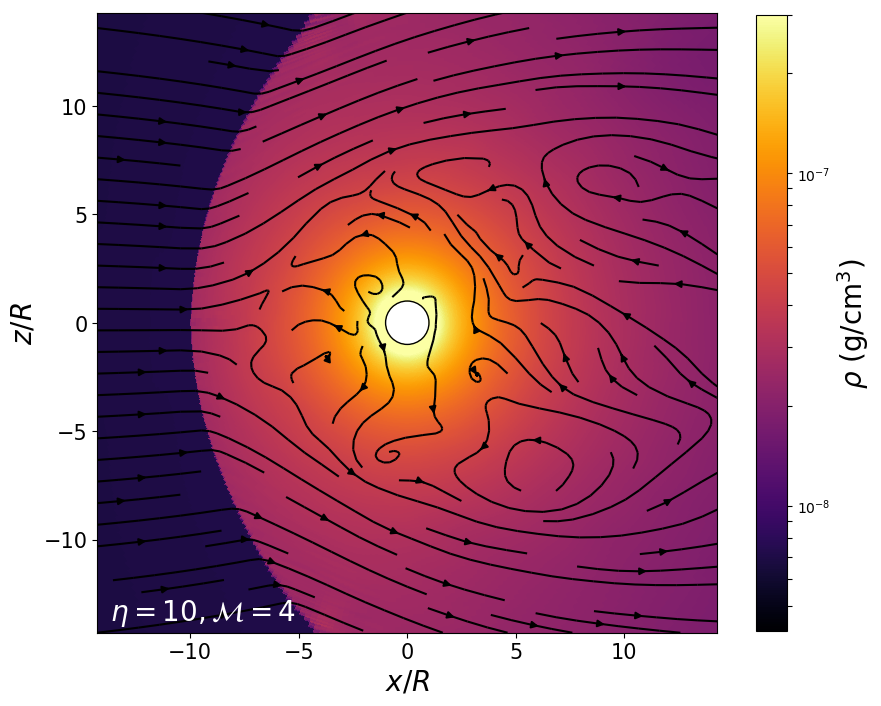}
    \caption{Density snapshots with velocity streamlines from our high-$\eta$ runs $\eta=4$ \textit{(top)}, $\eta=7$ \textit{(middle)} and $\eta=10$ \textit{(bottom)} with identical colormap limits.
    \label{fig:higheta}}
\end{figure}

\begin{figure}
  \includegraphics[width=0.5\textwidth]{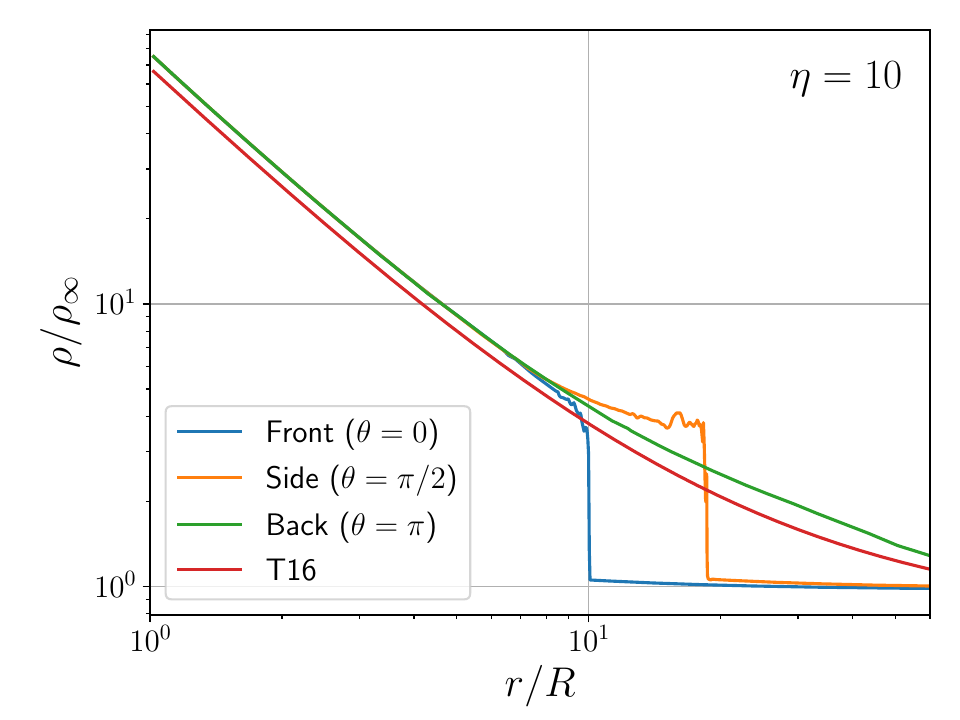}
  \includegraphics[width=0.5\textwidth]{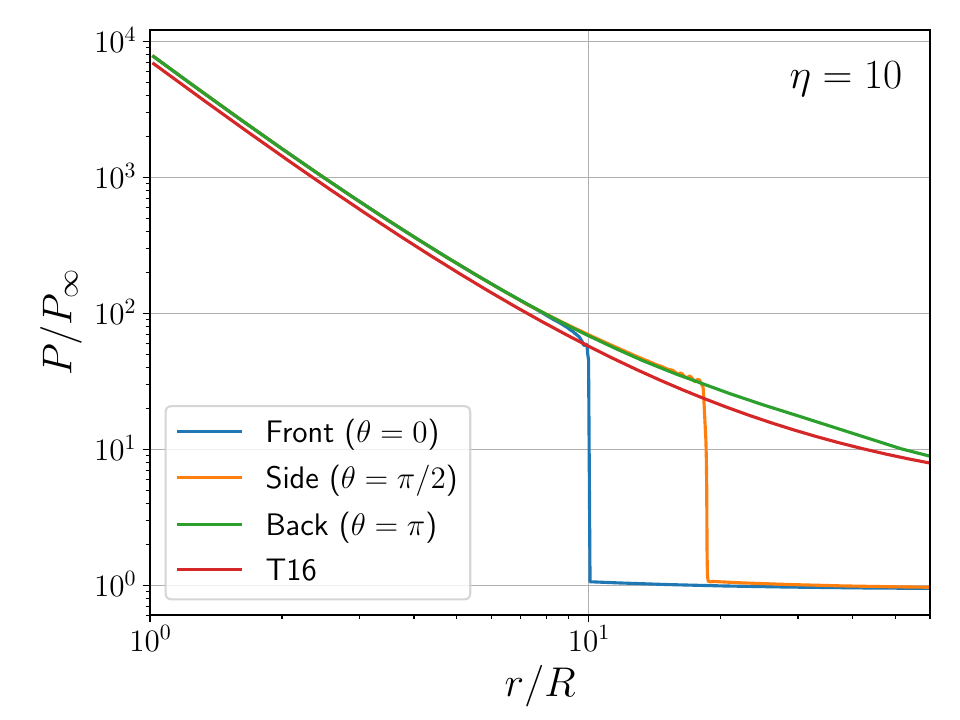}
  \includegraphics[width=0.5\textwidth]{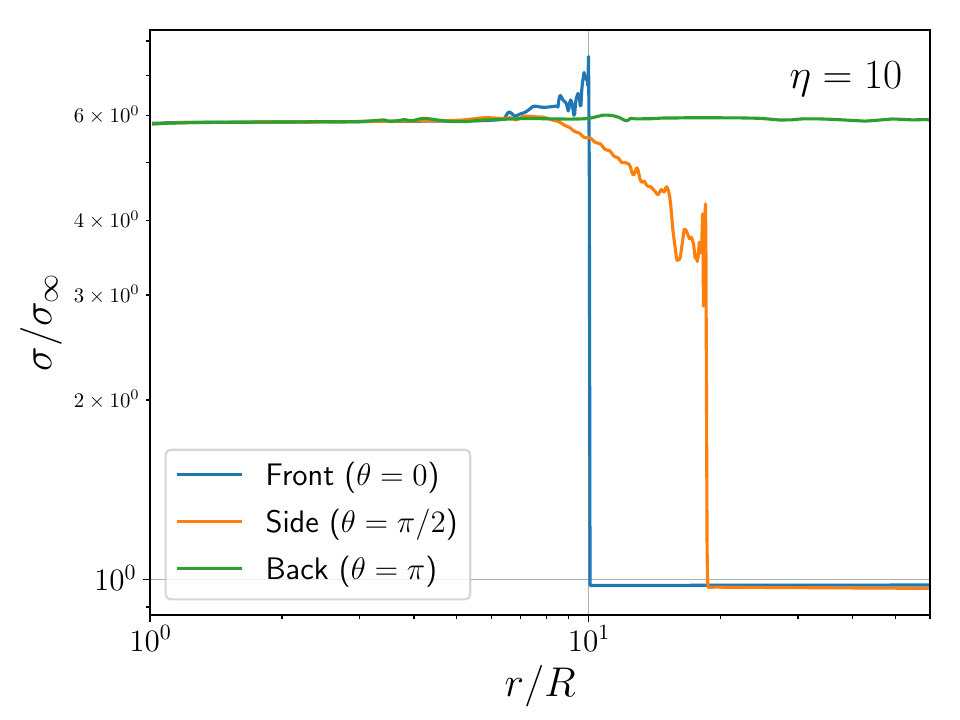}
    \caption{Density \textit{(top)}, pressure \textit{(middle)} and pseudoentropy \textit{(bottom)} profiles along radial rays from the centre of the sphere for $\eta=10$. Pressure and density predictions for the hydrostatic halo by T16 are also shown, which slightly underestimate the values we find here. Within the halo, the three radial slices are nearly identical, demonstrating spherical symmetry. The entropy fluctuates near the bow shock but is constant over much of the halo.
    \label{fig:profile}}
\end{figure}

\begin{figure}
  \includegraphics[width=0.5\textwidth]{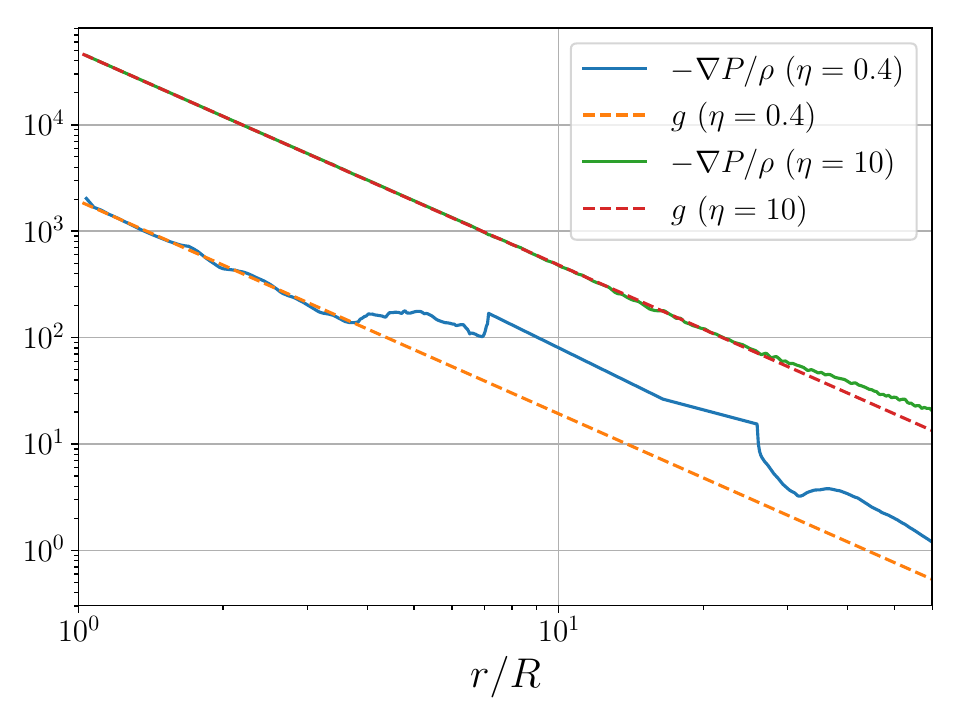}
    \caption{Pressure gradient and gravitational acceleration along a radial ray behind the sphere ($\theta=\pi$) for $\eta=0.4$, $\mach=4$ and $\eta=10$, $\mach=4$. Hydrostatic equilibrium is achieved when these quantities are equal.
    \label{fig:hseenv}}
\end{figure}

\begin{figure}
  \includegraphics[width=0.5\textwidth]{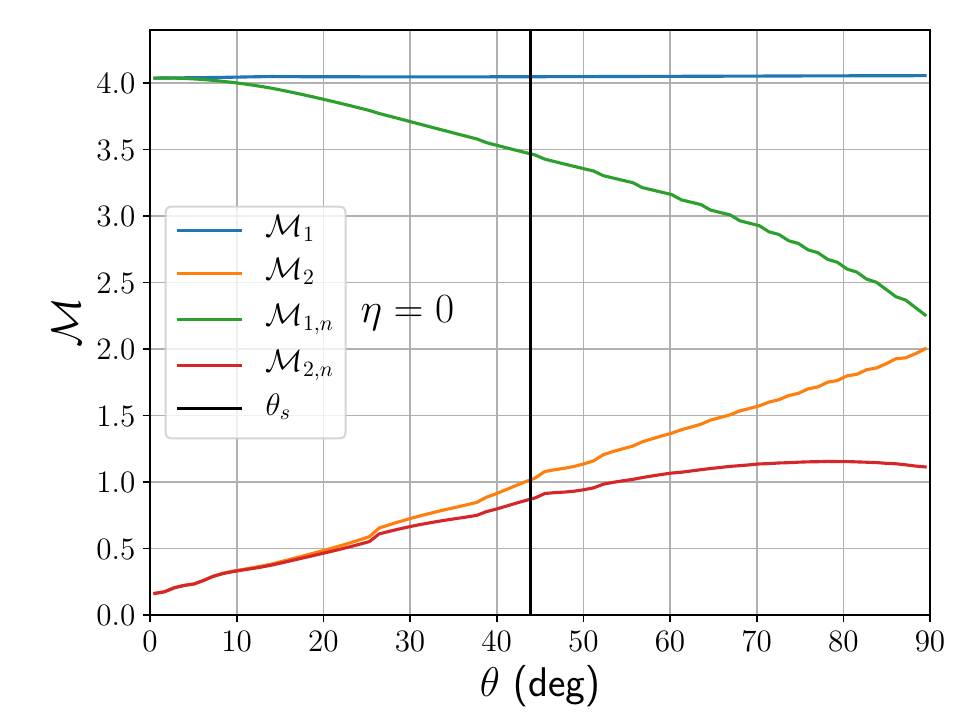}
  \includegraphics[width=0.5\textwidth]{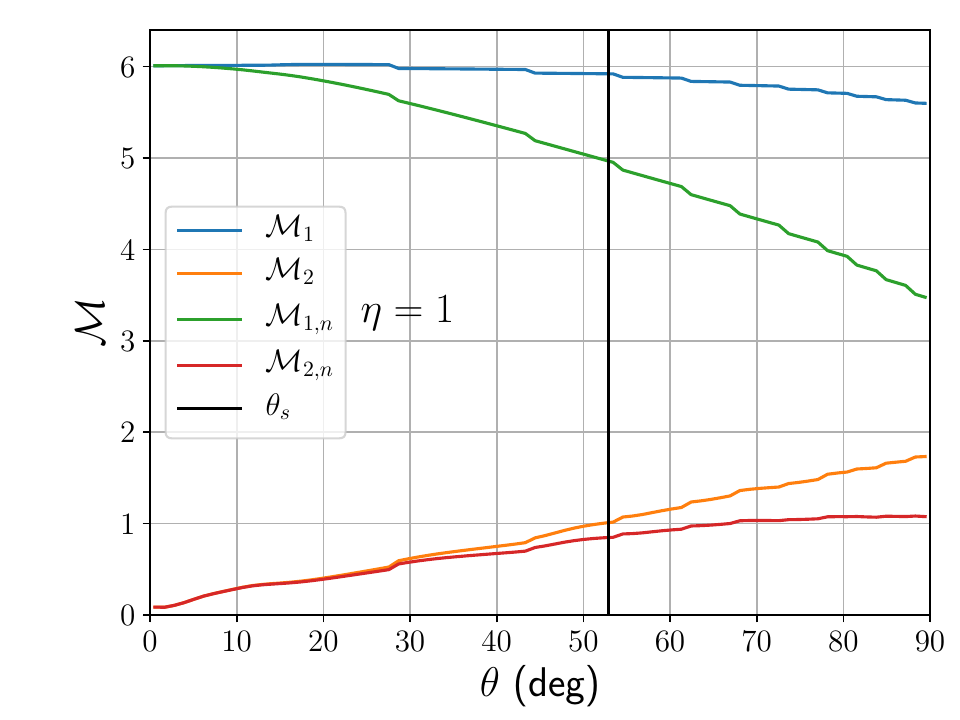}
  \includegraphics[width=0.5\textwidth]{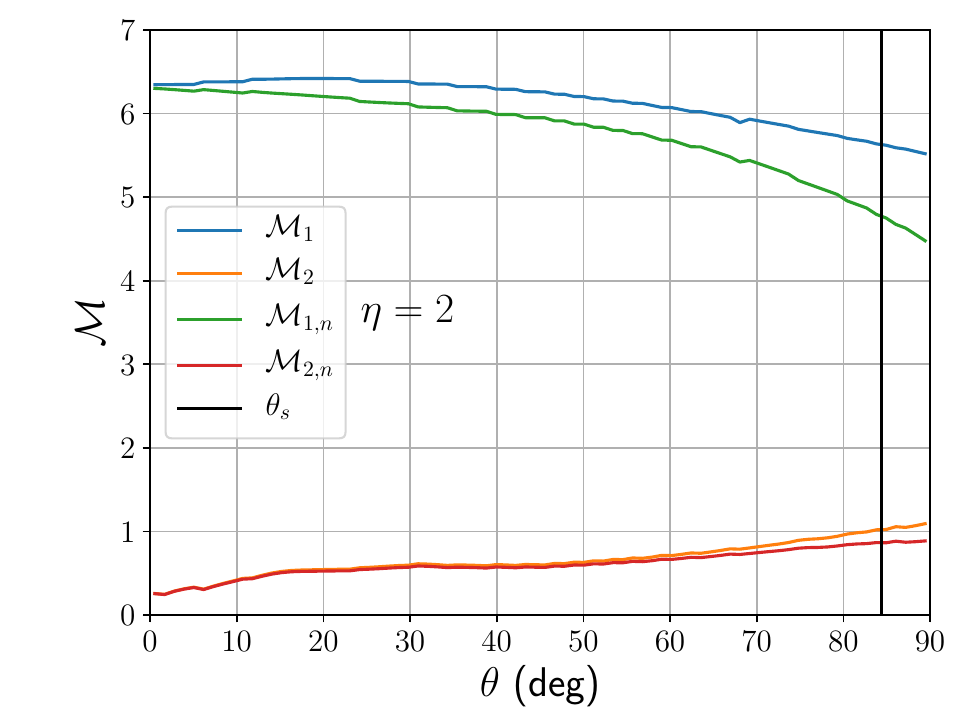}
    \caption{Mach number $\mach$ and its component normal to the bow shock $\mach_{n}$ immediately pre- and post-shock (subscripts 1 and 2, respectively) for $\eta=0$ \textit{(top)}, $\eta=1$ \textit{(middle)} and $\eta=2$ \textit{(bottom)} with $\mach_{\infty}=4$. The solid black line denotes the sonic line $\theta_{s}$, which moves to higher $\theta$ as $\eta$ is increased.
    \label{fig:mperp}}
\end{figure}

\begin{figure}
  \includegraphics[width=0.5\textwidth]{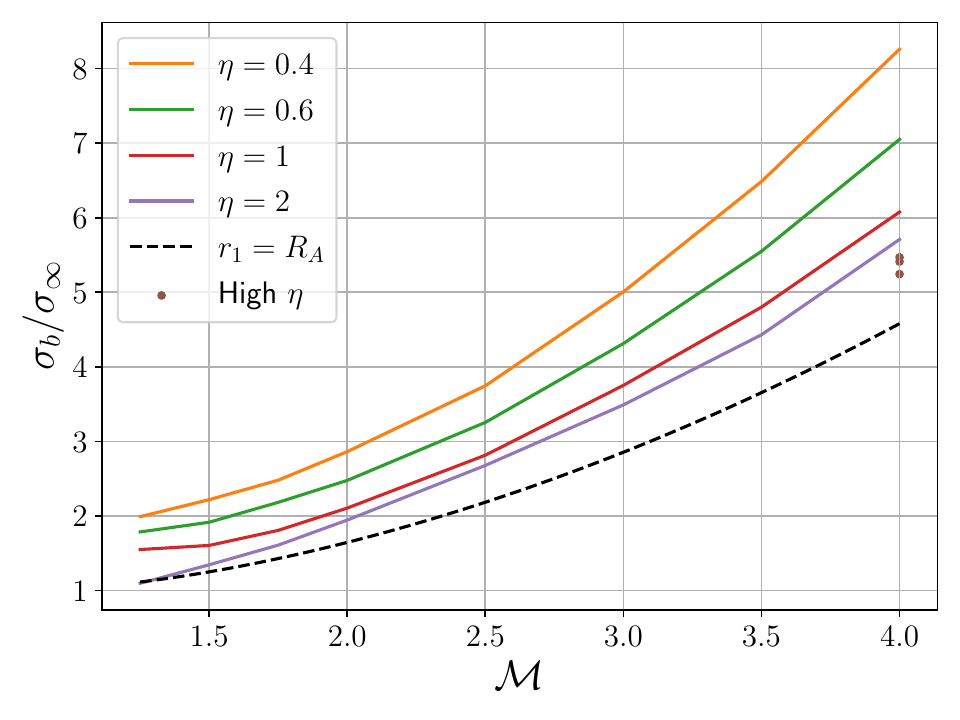}
    \caption{Pseudoentropy at the base of the hydrostatic halo or separation bubble. As $\eta$ increases, the entropy converges toward the lower bound set by the assumption that $r_{1}=\ra$ in conjunction with the jump conditions.
    \label{fig:entbubble}}
\end{figure}

\begin{table}
\begin{center}
\caption{Density and pressure at the base of the hydrostatic halo as well as the mass of the halo for our high-$\eta$ runs. The power-law scalings are also listed along with their expected values.}
\begin{tabular}{c c c c}
\hline
\hfill $\eta$ & $\rho_{b}/\rho_{\infty}$ & $P_{b}/P_{\infty}$ & $m_{\rm halo}\times 10^{6}M_{J}$ \\
\hline
\hfill 4  & 21.6 & 967 & 2.22 \\
\hfill 7  & 45.8 & 3400  & 12.2 \\
\hfill 10 & 75.0 & 7750  & 34.8 \\
\hline
\hfill $\eta$ scaling & 1.356 & 2.268 & 3.007 \\
\hfill predicted scaling & 1.5 & 2.5 & 3 \\
\hline
\label{tab:scaling}
\end{tabular}
\end{center}
\end{table}

Density snapshots from our calculations with $\eta>2$ are shown in Fig. \ref{fig:higheta}. Here, the gravity of the body is strong enough to cloak it in a halo of gas adorned with a single bow shock. The entropy and velocity gradients created by the bow shock have been shown to cause Rayleigh-Taylor and Kelvin-Helmholtz instabilities in the shocked material. This so-called ``flip-flop'' instability leads to oscillations in the surface of the shock cone \citep{1987MNRAS.226..785M,1999A&A...347..901F}. The instability is more violent for small and absorbing bodies and high Mach numbers \citep{1989MNRAS.236..817M,1991A&A...248..301M,1992MNRAS.255..183M,1997ApJ...478..723B,1998A&A...337..311S} and is suppressed in 3-D simulations compared to similar work in 2-D \citep{1994ApJ...427..351R}. A thorough survey of these instabilities and their relationship with dimensionality and numerical artefacts can be found in \citet{2005A&A...435..397F}. Indeed, eddies and vortices are visible in the low-speed flow within the halo in Fig. \ref{fig:higheta}, causing modest time variation in the flow even after a steady state has been reached. Because of this, we time-average our derived quantities for these runs.

Displayed in Fig. \ref{fig:profile} are radial slices of the density, pressure, and pseudoentropy for $\eta=10$. The nose of the bow shock is sharply defined in the $\theta=0$ slice. At a sufficient distance behind the shock, we see that the hydrodynamic variables in each of the three radial slices are identical, demonstrating spherical symmetry of the halo. T16 derived an analytical model for the radial profile of the halo in hydrostatic equilibrium (HSE) given the stand-off distance and post-shock fluid state (subscript 2): 
\be
\rho(r) = \rho_{2} \left[1-\frac{\gamma-1}{c_{s,2}^{2}/c_{s,\infty}^{2}}\frac{\mach_{\infty}^{2}}{2}\left(\frac{\ra}{\rs}-\frac{\ra}{r}\right)\right]^{1/(\gamma-1)}.
\ee
We recast this using $\rs=\eta R$, the definition of $\eta$ and thermodynamic relations as
\be
\rho(r) = \rho_{2} \left[1-\frac{G m}{h_{2}R}\left(\frac{1}{\eta}-\frac{R}{r}\right)\right]^{1/(\gamma-1)}. \label{eq:rhor2}
\ee
This prediction and its associated pressure slightly underestimate our pressure and density profiles (Fig. \ref{fig:profile}). The pressure and density at the surface of the sphere $r=R$, which may determined from (\ref{eq:rhor2}), far exceed those of the ambient medium. For $\eta \gg 1,\ G m \gg h_{2} R$, the scaling relation $\rho_{b} \propto \eta^{1/(\gamma-1)} = \eta^{3/2}$ holds at constant $\mach$. For pressure, we similarly obtain $P_{b} \propto \rho_{b}^{\gamma} \propto \eta^{5/2}$.

The total mass contained within this halo can be obtained by integration of (\ref{eq:rhor2}) over the halo, taking $\rs$ as its radial extent. We perform this integration in Appendix \ref{sec:halomass} and show that $m_{\rm halo}\propto\rho_{\infty}\eta^{3}$. In Table \ref{tab:scaling}, we list the values of $\rho_{b}$, $P_{b}$ and $m_{\rm halo}$ measured from our simulation output at high $\eta$. We also compute the power-law scaling for each variable and compare to the values predicted above. The halo mass scaling is in good agreement with our prediction, whereas the pressure and density scalings fall slightly short. This indicates that the assumptions $\eta \gg 1,\ G m \gg h_{2} R$ are not valid over this range of parameter space, which is unsurprising as the largest $\eta$ we have considered is 10. We expect these scaling relations to be obeyed at much larger values of $\eta$.

As mentioned in section \ref{sec:morphology}, the time required for these simulations to reach a steady state is approximately the fluid crossing time of the simulation domain $t_{cr}=2R_{\rm out}/v_{\infty}$. Another relevant timescale is that required to form a halo of mass $m_{\rm halo}$. Approximating the halo cross-section as $\pi\ra^2$, the mass flux feeding the halo is $\pi\ra^{2}\rho v_{\infty}$. Then the time needed to form the halo is
\be
t_{\rm halo} \approx \frac{m_{\rm halo}}{\pi\ra^{2}\rho v_{\infty}}.
\ee
Since $m_{\rm halo}$ scales as $\rho\ra^{3}$, this timescale goes as $t_{\rm halo}\propto\ra/v_{\infty}$. This is the fluid crossing time of the halo, which is clearly less than the crossing time of the simulation domain, so the limiting timescale in our simulations is $t_{cr}$.

In addition to the halo, HSE can also occur within the separation bubbles discussed in section \ref{sec:structure}. Such structures cannot be described by (\ref{eq:rhor2}), as the assumption underlying (\ref{eq:rhor2}) that the bow shock provides a well-defined outer boundary condition is no longer valid. Instead, we turn to the fundamental condition for HSE 
\be
\boldsymbol{g}=-\nabla P/\rho. \label{eq:hse}
\ee
The left and right sides of (\ref{eq:hse}) are computed from our simulation data and compared in Fig. \ref{fig:hseenv} for a separation bubble and a halo. We see that for $\eta=0.4$ the separation bubble is in hydrostatic equilibrium within several $R$ of the body, whereas for $\eta=10$ the equilibrium is maintained out to at least the stand-off distance $\rs=10R$.

Fig. \ref{fig:profile} also shows that the entropy within the halo is constant and uniform, varying at the percent level with the exception of some fluctuations near the shock front. This constant value is set by the entropy generation at the shock, \lp{located at $r_{1}(\theta)$,} which depends on the incoming Mach number at the bow shock $\mach_{1}$. From conservation of energy,
\be
\frac{1}{2} \mach_{\infty}^{2} = \frac{1}{2} \mach_{1}^{2} - \frac{G m}{c_{s}^{2} r_{1}}, \label{eq:machra}
\ee
assuming a constant sound speed prior to the shock. If one assumes that the shock is located at $r_{1} \approx R_{A}$ to obtain a lower bound, it follows that 
\be
\mach_{1}^{2} = 2\mach_{\infty}^{2}. \label{eq:m2m}
\ee
Values of $r_{1}$ greater than $R_{A}$ are also possible in the outer regions of the shock far from the body, but here the shock is weak and does not generate significant entropy. Furthermore, since the shock is oblique, the Mach number normal to the shock $\mach_{1,n}$ will be no greater than that predicted by (\ref{eq:m2m}). It is not possible to analytically determine $\mach_{1,n}$, but we show our numerical results in Fig. \ref{fig:mperp}. We determine $\mach_{n}=\mach\sin\beta$ by fitting the shape of the shock to determine its angle relative to the local velocity field $\beta$. Without gravity, $\beta$ is equal to the local shock angle and varies between $\pi$ and the angle of the Mach cone $\mu=\arccos(\sqrt{\mach^{2}-1}/\mach)$. The presence of gravity alleviates this variation somewhat by turning the streamlines normal to the shock. We find that $\mach_{1,n}$ falls off more slowly with $\theta$ as gravity is increased. We also see that the position of the sonic line $\theta_{s}$ increases with mass. Thus, for increasing $\ra$ the strength of the bow shock and the entropy generated by that shock become more uniform.

Once $\mach_{1}$ has been determined, the entropy ratio can be found using (\ref{eq:m2m}) and the Rankine-Hugoniot shock jump conditions
\be
\frac{\rho_{2}}{\rho_{1}} &=& \frac{(\gamma+1)\mach_{1}^{2}}{(\gamma-1)\mach_{1}^{2}+2} = \frac{(\gamma+1)\mach_{\infty}^{2}}{(\gamma-1)\mach_{\infty}^{2}+1}, \\
\frac{P_{2}}{P_{1}} &=& \frac{2\gamma\mach_{1}^{2}-(\gamma-1)}{\gamma+1} = \frac{4\gamma\mach_{\infty}^{2}-(\gamma-1)}{\gamma+1},
\ee
where the subscript 2 denotes post-shock conditions. The pseudoentropy ratio is then
\be
\frac{\sigma_{2}}{\sigma_{1}} = \frac{P_{2}}{P_{1}} \left(\frac{\rho_{2}}{\rho_{1}}\right)^{-\gamma}. \label{eq:entratio}
\ee
We can use our lower bound on $\mach_{1}$ (\ref{eq:m2m}) in conjunction with the jump conditions to set a lower bound on the entropy. For $\mach_{\infty}=4$, we predict that
\be
\frac{\sigma_{2}}{\sigma_{1}} \geq 4.58.
\ee
From Fig. \ref{fig:profile}, we see that the numerical value is $\approx$6, which falls above our bound. This bound can be applied to the rest of our runs as well, as shown in Fig. \ref{fig:entbubble}. For our low-$\eta$ runs, which do not have a hydrostatic halo, we take the entropy at the base of the separation bubble. We see that the entropy converges toward our lower bound as $\eta$ is increased.

\section{Drag Forces} \label{sec:drag}

Both gravitational and hydrodynamic forces are at work on the body. The overdense wake and Mach cone trailing the sphere lead to a net gravitational force, known as dynamical friction. The non-uniform pressure distribution over the surface of the body owing to the ram pressure, shock structure and possible vortices also provides a net force.

\subsection{Calculating Dynamical Friction} \label{sec:df}

\begin{figure}
  \includegraphics[width=0.5\textwidth]{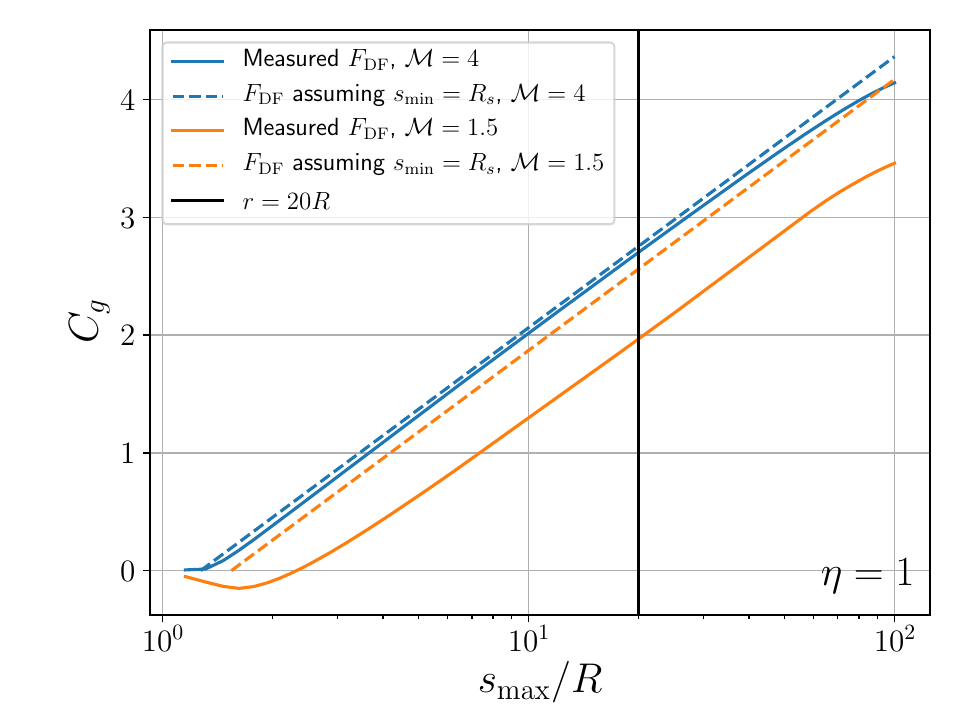}
  \includegraphics[width=0.5\textwidth]{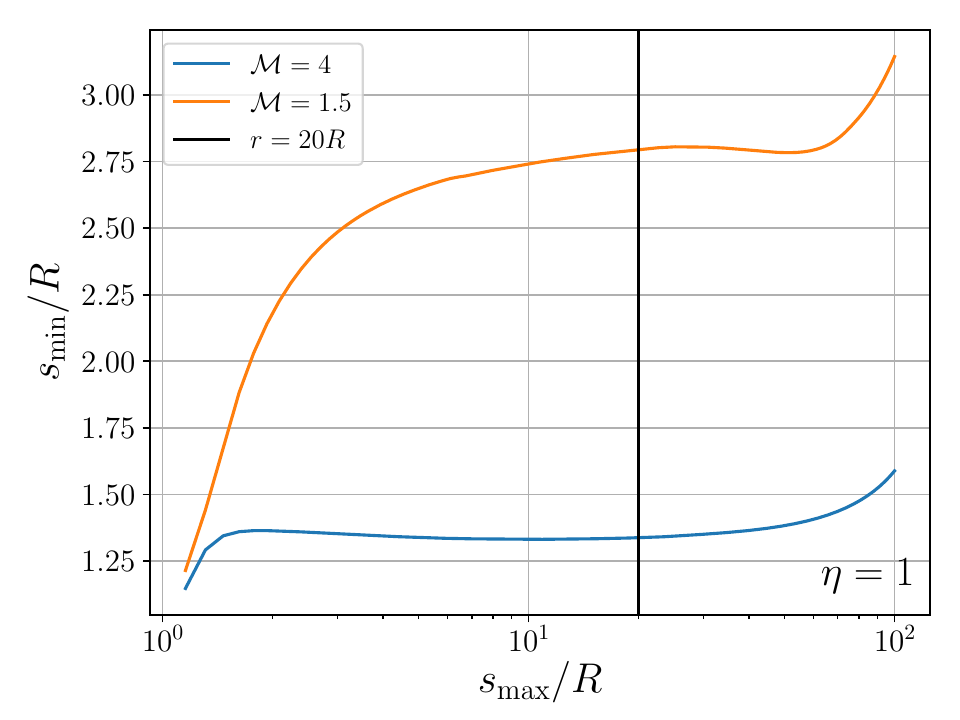}
    \caption{Gravitational drag coefficient $C_{g}$ \textit{(top)} and $\smin$ \textit{(bottom)} as a function of wake extent $\smax$ for $\eta=1$ with $\mach=4$ (blue lines) and $\mach=1.5$ (orange lines). For intermediate radii, the slope of $C_{g}$ agrees well with theoretical predictions so that $\smin$ plateaus. We choose to measure $\smin$ at $r=20R$ (black vertical line), as this typically falls within the plateau.
    \label{fig:dfrad}}
\end{figure}

\begin{figure}
  \includegraphics[width=0.5\textwidth]{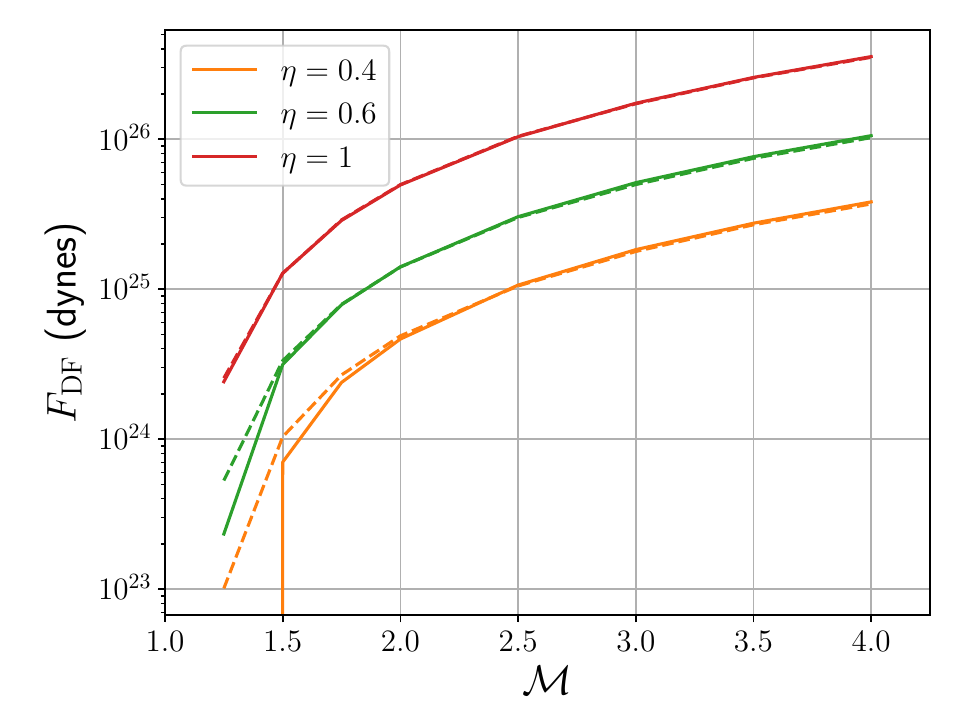}
  \includegraphics[width=0.5\textwidth]{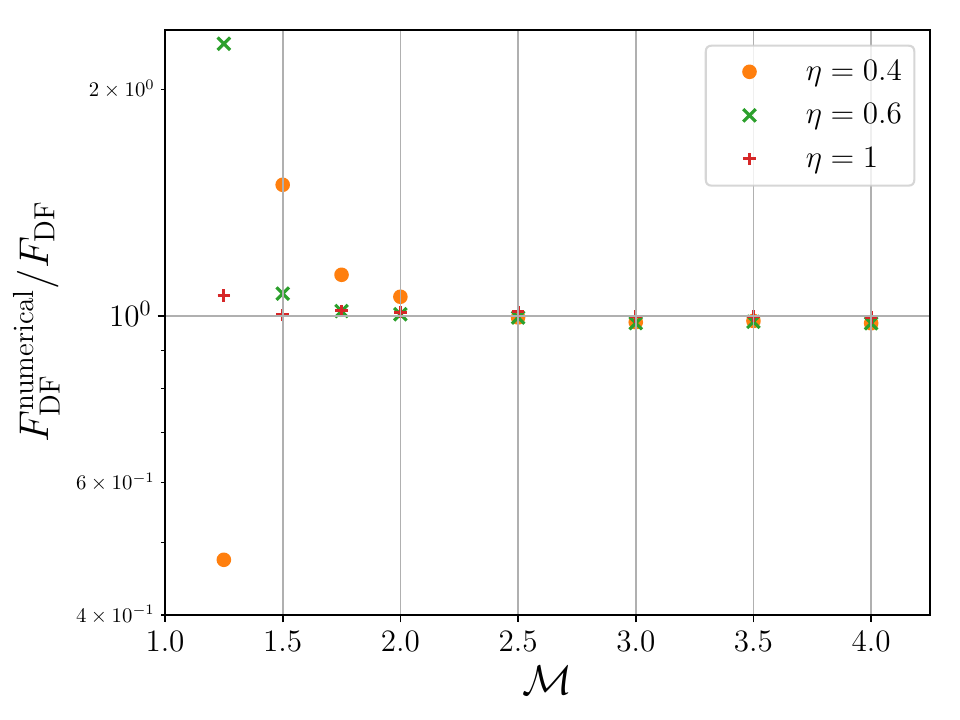}
    \caption{\lp{\textit{(Top)}} Comparison of the dynamical friction measured at $r=20R$ \lp{(solid lines)} against the theoretical value using our definition of $s_{\rm min}$ \lp{(dotted lines).} \lp{\sout{\textit{(top)} and the ratio between the two \textit{(bottom)}.}} \lp{\textit{(Bottom)} The ratio between the numerical and theoretical values.}
    \label{fig:dfsmin}}
\end{figure}

The dynamical friction on a supersonic body can be expressed as
\be
\fdf = 4\pi\rho_{\infty} \left(\frac{G m}{v_{\infty}}\right)^{2} \ln\left(\frac{\smax}{\smin}\right), \label{eq:fdf}
\ee
\citep{1971ApJ...165....1R}, where the Coulomb logarithm depends on the effective linear size of the body $\smin$ and the maximum extent of the wake $\smax$. For an infinite uniform medium, the extent of the wake is $\smax=v_{\infty}t$, so (\ref{eq:fdf}) predicts that $\smax\rightarrow\infty$ and $\fdf\rightarrow\infty$ as $t\rightarrow\infty$ \citep{1999ApJ...513..252O}. This is unphysical, as the increasing dynamical friction will slow the object into the subsonic regime where (\ref{eq:fdf}) is no longer valid. Thus, there is no steady-state solution for dynamical friction in an infinite medium. In a planetary engulfment, the size of the giant provides an upper bound for the extent of the wake, but $\smax$ remains largely unconstrained and is often approximated as the orbital separation.

The choice of $\smin$ is ambiguous, with authors finding a variety of fits \citep{1999ApJ...522L..35S,2009ApJ...703.1278K,2011MNRAS.418.1238C,2013ApJ...775...72B,2021arXiv210315848M}. However, these works use either a point mass or a Plummer potential
\be
\Phi_{P}=-\frac{G m}{\sqrt{r^{2}+r_{s}^{2}}}, \label{eq:plummer}
\ee
and take the softening length $r_{s}$ as a measure of the size of of the engulfed body. It is not clear how $r_{s}$ relates to the physical radius $R$ of a solid sphere and whether the scaling relations found for a Plummer potential also hold for a solid body, particularly when the shock is near the surface of the body. T16 showed that the stand-off distance $\rs$ is an appropriate choice for $\smin$ in the range $1<\eta<50$, but that this choice overestimates the dynamical friction for $\eta\leq1$ (see their Fig. 16).

Due to our use of a uniform medium, we are ill-equipped to inform the choice of $\smax$, but we can determine the effective linear size $\smin$. We measure the dynamical friction from our simulation output as
\be
\fdf = \int_{R}^{\smax} \frac{G m \rho}{r^{2}} \frac{\meshr\cdot\vel_{\infty}}{r v_{\infty}} dV = G m \sum_{r_{i}<\smax} \frac{\rho_{i} V_{i}}{r_{i}^{2}} \cos\theta_{i},
\ee
where $V_{i}$ is the volume of the $i$-th cell. This can also be expressed as a drag coefficient via 
\be
\fdf = \pi \ra^{2} \rho v_{\infty}^{2} C_{g} = 4 \pi \rho \left(\frac{G m}{v_{\infty}}\right)^{2} C_{g},
\ee
so that
\be
C_{g} \equiv \ln\left(\frac{\smax}{\smin}\right) = \ln\smax-\ln\smin. \label{eq:cg}
\ee
We show $C_{g}$ in the top panel of Fig. \ref{fig:dfrad} for $R<\smax<R_{\rm out}$ (solid lines) as well as the prediction of T16 (dotted lines), which asserts that $\smin=\rs$.
We find that $C_{g} \propto \smax$, as expected, but that for low $\eta$ and $\mach$ the curves are offset from their predicted values. Note from (\ref{eq:cg}) that $\ln(\smin)$ sets the ``intercept'' of the curve. Therefore, an offset between the two curves indicates that $\smin$ differs from $\rs$. Near the outer boundary, the slope of $C_{g}$ vs $\ln(\smax)$ drops below $4\pi\rho_{\infty}(G m/v_{\infty})^{2}$, the value expected from (\ref{eq:cg}). We find that this discrepancy is attributable to the finite size of the domain, as variation of the domain size $R_{\rm out}$ reveals that this feature always occurs near $r=R_{\rm out}$. This and other effects of the finite domain are discussed in Appendix \ref{sec:convergence}.

For a given $\smax$, we can determine $\smin$ by inverting (\ref{eq:fdf}):
\be
\smin = \smax \exp\left[-\frac{\fdf}{4\pi\rho_{\infty}}\left(\frac{v_{\infty}}{G m}\right)^{2}\right],
\ee
We see that $\smin$ is fairly uniform throughout the domain except for $r\approx R$ and in the outer regions of the domain (bottom panel of Fig. \ref{fig:dfrad}). This excess in $\smin$ at large $\smax$ is unphysical and is due to the finite domain size, as mentioned above. Thus, to obtain an accurate $\smin$ we need only choose $\smax$ significantly greater than $\rs$ and well within $R_{\rm out}$. For consistency, we choose $\smax=20R$, which lies within the plateau evident in the bottom panel of Fig. \ref{fig:dfrad}. We find that $\smin$ can be well-approximated by the fit $\smin/\rs = 2R/\ra$, which can be rewritten in the alternative forms
\be
\frac{\smin}{\rs} = 2\frac{R}{\ra} = \frac{1}{\eta}\frac{\mach^{2}}{\mach^{2}-1} = \frac{1}{\eta\cos^{2}\mu}, \label{eq:smin}
\ee
where $\mu=\arccos(\sqrt{\mach^{2}-1}/\mach)$ is the Mach angle. We check this result by determining $\fdf$ using (\ref{eq:fdf}), where we plug in (\ref{eq:smin}) and $\smax=20R$. Here the shock stand-off distance is the only quantity we need enter ``by hand.'' This prediction is shown in the top panel of Fig. \ref{fig:dfsmin} (dotted lines) along with the values measured from our simulations (solid lines). We also show the ratio between our predictions and measurements in the bottom panel. We find agreement between our predictions and measurements, particularly for $\eta=1$. To obtain an estimate of the dynamical friction across all $\eta$, we combine our results with those of T16 for $\eta>1$:
\be
\frac{\smin}{R_{s}} = \begin{cases} 2\frac{R}{R_{A}} & \eta \leq 1, \\ 1 & \eta > 1. \end{cases} \label{eq:smincombined}
\ee
At $\eta=1$, these predictions differ by a factor of $\mach^{2}/(\mach^{2}-1)$, which is near unity for sufficiently large $\mach$.

$C_{g}$ is often negative within a few $R$ of the sphere, but far from the body is always positive and obeys the expected $C_{g}\propto\ln(\smax)$ behaviour. Indeed, the point $\eta=0.4$, $\mach=1.25$ is an outlier in Fig. \ref{fig:dfsmin} because the dynamical friction is negative even at $\smax=20R$. Thus, the choice of $\smax$ determines not only the magnitude but also the sign of $\fdf$.

\citet{1980ApJ...240...20R} showed that the dynamical friction on a supersonic body contains another term in addition to the Coulomb logarithm which becomes large near $\mach=1$:
\be
\fdf = 4\pi\rho_{\infty} \left(\frac{G m}{v_{\infty}}\right)^{2} \left[\ln\left(\frac{\smax}{\smin}\right) - \frac{1}{2}\ln(1-\mach^{-2})\right].
\ee
If this term is included, we need only modify our formula (\ref{eq:smincombined}) via the substitution
\be
\smin \longrightarrow \smin \sqrt{1-\mach^{-2}}.
\ee

\subsection{Pressure Drag} \label{sec:pressure}

\begin{figure}
  \includegraphics[width=0.5\textwidth]{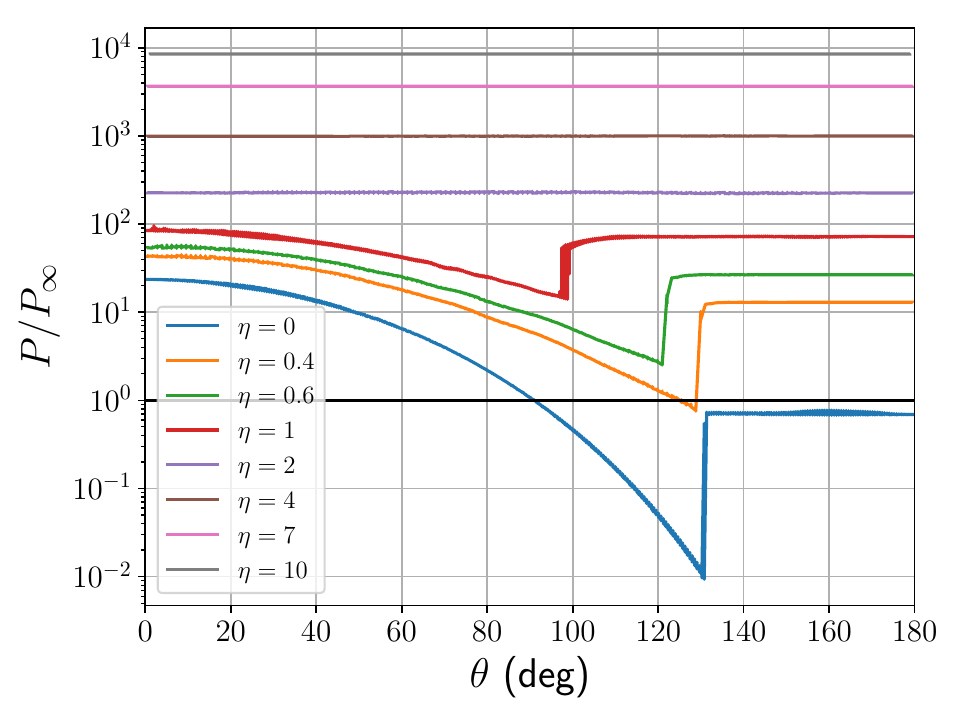}
    \caption{Pressure distribution over the surface of the engulfed body at $\mach=4$ for all $\eta$. The pressure is uniform for high $\eta$, while the separation bubble is clearly defined for low $\eta$.
    \label{fig:presdist}}
\end{figure}

\begin{figure}
  \includegraphics[width=0.5\textwidth]{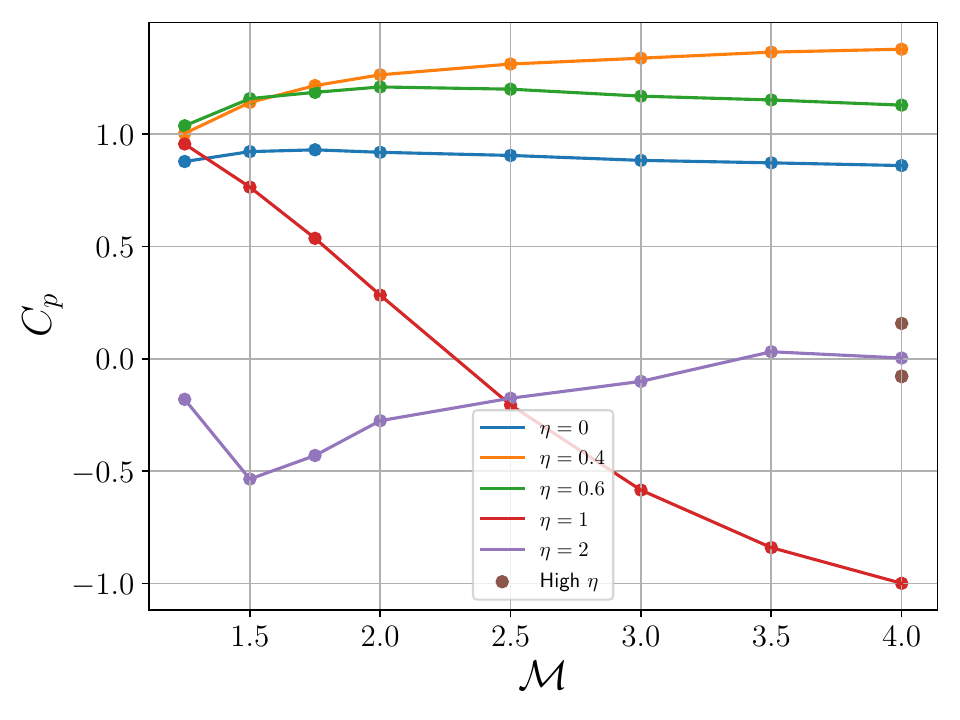}
  \includegraphics[width=0.5\textwidth]{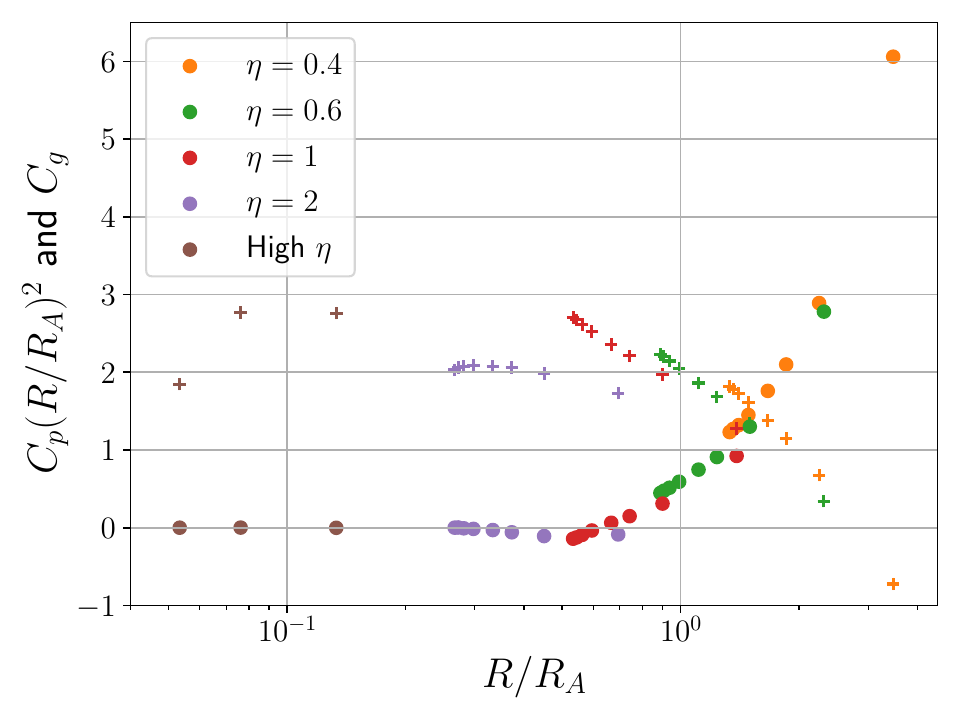}
    \caption{\textit{(Top)} Coefficient of pressure drag as a function of Mach number. For $\eta<1$, $C_{p}$ remains near unity, while for $\eta=1$ it varies from 1 down to $-1$. At higher $\eta$, $C_{p}$ tends toward zero and may be slightly positive or negative. \textit{(Bottom)} Gravitational drag coefficient ($+$ signs) and pressure drag coefficient multiplied by $(R/R_{A})^{2}$ (circles) for the purpose of comparison. Simulations with $\eta=0$ are excluded from the bottom panel since they have $R/R_{A}=\infty$. For large $R_{A}$, $C_{p}(R/R_{A})^{2}\rightarrow0$ as $R_{A}\rightarrow\infty$, as expected. For moderate $R_{A}$, we find slightly negative values, though they are much smaller than $C_{g}$ regardless.
    \label{fig:cp}}
\end{figure}

An engulfed body also experiences hydrodynamic drag as a result of a non-uniform pressure distribution over its surface. The pressure distribution for our $\mach=4$ runs is shown in Fig. \ref{fig:presdist}. The pressure is uniform for high $\eta$, while for low $\eta$ the boundary of the separation bubble is sharply defined. The force provided by the bubble can rival or even (for $\eta=1$) exceed that on the front of the sphere. The pressure drag can be expressed as 
\be
F_{p} = \frac{1}{2} \pi R^{2} \rho v^{2} C_{p}, \label{eq:fp}
\ee
for some pressure drag coefficient $C_{p}$. The dependence of $C_{p}$ on Mach number and Reynolds number at 
very high Reynolds number is not precisely known, but is usually quoted as anywhere from 0.5 to 1 \citep{1972AIAAJ..10.1436B}. We measure $F_{p}$ from our simulation output via the surface integral of the pressure distribution
\be
F_{p} = \int_{S} P dS \cos\theta = R^{2} \sum_{r_{i}=R} P_{i} \Delta\theta_{i} \Delta\phi_{i} \sin\theta_{i} \cos\theta_{i}.
\ee
Then (\ref{eq:fp}) can be used to calculate $C_{p}$, which we plot in the top panel of Fig. \ref{fig:cp}. We find that $C_{p}$ is of order unity for $\eta<1$, while for $\eta=1$ it transitions from 1 all the way to $-1$. For $\eta=2$ the drag coefficient goes to zero as $\ra$ is increased, and is near zero for larger values of $\eta$. This is unsurprising due to the spherically-symmetric halo at high $\eta$, though persistent fluctuations in the flow due to instabilities cause $C_{p}$ to oscillate around zero.

For the purposes of comparison with the dynamical friction, \citet{yarza} plotted the quantity $C_{p}(R/\ra)^{2}$ against $R/\ra$, as the prefactors in (\ref{eq:fp}) and (\ref{eq:fdf}) differ by a factor of $(R/\ra)^{2}$. We follow suit in the bottom panel of Fig. \ref{fig:cp}, including $C_{g}$ for comparison, though we emphasize that the behaviour of $C_{g}$ depends on the choice of $\smax$ which we have chosen arbitrarily. We find slightly negative values of $C_{p}$ over a range of $R/\ra$, though dynamical friction dominates the total drag in this regime regardless. These results indicate that the standard form for the hydrodynamical drag (\ref{eq:fp}) with a drag coefficient near unity is sufficient for the regime in which $F_{p}$ is significant.

\subsection{Domain Size} \label{sec:domain}

In Appendix \ref{sec:convergence}, we perform a convergence test in which the domain size $R_{\rm out}$ is varied. This is necessary in the presence of a gravitating body, as the upstream flow is deflected by gravity in the far field. Thus, it is necessary to use a domain large enough that our assumption that incoming streamlines are parallel along the $\hat{x}$-direction is valid. The results of this test are listed in Table \ref{tab:setups}, where the drag coefficients are shown to vary by $\sim$50 per cent over our parameter space, converging at $R_{\rm out}\approx50\ra$. We conclude that this is the minimum domain size needed to obtain accurate measurements of the drag coefficients in the presence of gravity.

\section{Discussion and Conclusions} \label{sec:discussion}

In this paper we investigated the morphology of the flow around a rigid supersonic gravitating sphere engulfed in a homogeneous gaseous medium using 3-D hydrodynamical simulations in \athena. We focused our parameter space and analysis on the case of a planet engulfed by a giant star, though our results are applicable to any problem with a range of Mach numbers and $\ra/R$ comparable to what we consider here. In particular, we focus on the $\eta\leq1$ regime, which has received relatively little study. We also extend the supersonic results of T16 to lower $\mach$, as engulfed planets are expected to have Mach numbers near unity over a large portion of the stellar envelope. The shock stand-off distance is found to agree with T16 at high Mach number, but departs significantly as $\mach$ approaches unity.

We find that there are two distinct regimes of flow morphology. At low $\eta$, the morphology resembles that of a non-gravitating sphere: a bow shock forms near the surface of the body, while downstream a recompression shock is provided by flow separation from the surface. However, the flow departs from the non-gravitating result due to gravitational acceleration and deflection of the flow upsteam of the bow shock. Furthermore, the base of the separation bubble is hydrostatically-supported and the bubble grows in size as the mass of the gravitating body is increased, though it is not entirely clear if this growth of the bubble is due to its own hydrostatic support or to changes in morphology upstream. The transition between these two morphologies occurs at $\eta\approx1$. Here the wake grows large enough that the recompression shock intersects with the bow shock, and a further increase in $\eta$ causes these two shocks to combine into a single bow shock. Based on the range of $\eta$ over the inspirals computed by \citet{2023ApJ...950..128O}, this transition is expected to occur within the stellar envelope during a planetary engulfment event.

For high $\eta$, the sphere is cloaked in a spherically-symmetric hydrostatic halo with a stand-off distance described by (\ref{eq:rsthun}) and density profile (\ref{eq:rhor2}), as shown by T16. The entropy of the halo is set at the shock front, and is fairly uniform due to the deflection of the streamlines by gravity prior to reaching the shock. The entropy gradients which remain are near the bow shock and produce instabilities which cause oscillations in the shock cone, but are suppressed in 3-D and for the relatively small values ($<20$) of $\ra/R$ we consider here \citep{1994ApJ...427..351R,2005A&A...435..397F}. The hydrostatic profile of the halo (\ref{eq:rhor2}) can be determined by setting an outer boundary condition at the stand-off distance using the jump conditions. The fluid state at the centre of the halo (i.e. at the surface of the body) as well as the halo mass can be determined from this profile, and scale as powers of $\eta$.

These morphologies have consequences for the drag forces experienced by the body. For the dynamical friction, the two key unknowns are the maximum extent of the wake and the effective linear size of the perturber. We find that the effective linear size can be approximated by the ratio of the stand-off distance to half the accretion radius for $\eta\leq1$. The theoretical underpinnings of this result are not clear, though it performs well for $\eta\leq1$. When $\eta>1$, we defer to the result of T16 that the effective linear size is equal to the stand-off distance. Together, these results offer a complete model of dynamical friction for a supersonic gravitating sphere (\ref{eq:smincombined}).

The pressure drag is similarly bifurcated: in the case of a hydrostatic halo, the spherical symmetry of the fluid near the body results in a small pressure drag. This compounded by the fact that such objects have high dynamical friction, so we conclude that the pressure drag is negligible for $\eta>1$. For low $\eta$, the pressure drag is well-approximated by the standard non-gravitating case (\ref{eq:fp}) with a drag coefficient near unity. This is in agreement with studies on the migration of planetesimals by \citet{2015ApJ...811...54G}.

Finally, we investigate the effects of the simulation domain size on our results. Because we assume that the incoming streamlines are parallel, the use of a small domain risks underestimating the deflection of the streamlines due to gravity which can cause errors of $\sim$50 per cent in the drag coefficients. We find that a domain size comparable to $\sim$$50R_{A}$ is necessary for convergence. We have assumed that the accretion radius is small compared to the length scale of any gradients in the medium (e.g. the local scale height of a stellar profile) so that we can assume a uniform medium, though this may not be valid for massive planets or small orbital separations in a planetary engulfment event.

We have not considered subsonic velocities, as they are not well-suited to our methods. There is no \textit{a priori} reason why subsonic velocities cannot occur in any of the physical scenarios we have described, so future study on this regime is needed for a complete understanding of flow around a gravitating sphere.

\section*{Data Availability}

The fork of \athena\ used to perform the simulations described here can be found at \url{https://github.com/ljprust/athena/tree/windtunnel}, configured with the problem generator src/pgen/windtunnel.cpp. The parameter file for the fiducial case $\eta=1$, $\mach=4$ is located at inputs/hydro/athinput.windtunnel\textunderscore sph. This parameter file can be modified according to Table \ref{tab:setups} to reproduce any result presented in this paper.

\section*{Acknowledgements}

We thank Christopher O'Connor and Philip Chang for numerous and useful discussions. This research was supported in part by the NSF under Grant No. NSF PHY-1748958 and by a grant from the Simons Foundation (216179, LB). Computational resources for the calculations performed in this work were provided by the Mortimer HPC System, funded by NSF Campus Cyberinfrastructure Award OAC-2126229 and the University of Wisconsin-Milwaukee. We use the {\sevensize MATPLOTLIB} \citep{Hunter:2007} and {\sevensize SCIPY} \citep{2020SciPy-NMeth} software packages for the generation of plots in this paper.




\bibliographystyle{mnras}
\bibliography{references}

\begin{thebibliography}{}
\makeatletter
\relax
\def\mn@urlcharsother{\let\do\@makeother \do\$\do\&\do\#\do\^\do\_\do\%\do\~}
\def\mn@doi{\begingroup\mn@urlcharsother \@ifnextchar [ {\mn@doi@}
  {\mn@doi@[]}}
\def\mn@doi@[#1]#2{\def\@tempa{#1}\ifx\@tempa\@empty \href
  {http://dx.doi.org/#2} {doi:#2}\else \href {http://dx.doi.org/#2} {#1}\fi
  \endgroup}
\def\mn@eprint#1#2{\mn@eprint@#1:#2::\@nil}
\def\mn@eprint@arXiv#1{\href {http://arxiv.org/abs/#1} {{\tt arXiv:#1}}}
\def\mn@eprint@dblp#1{\href {http://dblp.uni-trier.de/rec/bibtex/#1.xml}
  {dblp:#1}}
\def\mn@eprint@#1:#2:#3:#4\@nil{\def\@tempa {#1}\def\@tempb {#2}\def\@tempc
  {#3}\ifx \@tempc \@empty \let \@tempc \@tempb \let \@tempb \@tempa \fi \ifx
  \@tempb \@empty \def\@tempb {arXiv}\fi \@ifundefined
  {mn@eprint@\@tempb}{\@tempb:\@tempc}{\expandafter \expandafter \csname
  mn@eprint@\@tempb\endcsname \expandafter{\@tempc}}}

\bibitem[\protect\citeauthoryear{{Bailey} \& {Hiatt}}{{Bailey} \&
  {Hiatt}}{1972}]{1972AIAAJ..10.1436B}
{Bailey} A.~B.,  {Hiatt} J.,  1972, \mn@doi [AIAA Journal] {10.2514/3.50387},
  \href {https://ui.adsabs.harvard.edu/abs/1972AIAAJ..10.1436B} {10, 1436}

\bibitem[\protect\citeauthoryear{{Bailey}, {Stone}  \& {Fung}}{{Bailey}
  et~al.}{2021}]{2021ApJ...915..113B}
{Bailey} A.,  {Stone} J.~M.,   {Fung} J.,  2021, \mn@doi [\apj]
  {10.3847/1538-4357/ac033b}, \href
  {https://ui.adsabs.harvard.edu/abs/2021ApJ...915..113B} {915, 113}

\bibitem[\protect\citeauthoryear{{Bauer}, {White}  \& {Bildsten}}{{Bauer}
  et~al.}{2019}]{2019ApJ...887...68B}
{Bauer} E.~B.,  {White} C.~J.,   {Bildsten} L.,  2019, \mn@doi [\apj]
  {10.3847/1538-4357/ab4ea4}, \href
  {https://ui.adsabs.harvard.edu/abs/2019ApJ...887...68B} {887, 68}

\bibitem[\protect\citeauthoryear{{Benensohn}, {Lamb}  \& {Taam}}{{Benensohn}
  et~al.}{1997}]{1997ApJ...478..723B}
{Benensohn} J.~S.,  {Lamb} D.~Q.,   {Taam} R.~E.,  1997, \mn@doi [\apj]
  {10.1086/303835}, \href
  {https://ui.adsabs.harvard.edu/abs/1997ApJ...478..723B} {478, 723}

\bibitem[\protect\citeauthoryear{{Bernal} \& {S{\'a}nchez-Salcedo}}{{Bernal} \&
  {S{\'a}nchez-Salcedo}}{2013}]{2013ApJ...775...72B}
{Bernal} C.~G.,  {S{\'a}nchez-Salcedo} F.~J.,  2013, \mn@doi [\apj]
  {10.1088/0004-637X/775/1/72}, \href
  {https://ui.adsabs.harvard.edu/abs/2013ApJ...775...72B} {775, 72}

\bibitem[\protect\citeauthoryear{{Billig}}{{Billig}}{1967}]{1967JSpRo...4..822B}
{Billig} F.~S.,  1967, \mn@doi [Journal of Spacecraft and Rockets]
  {10.2514/3.28969}, \href
  {https://ui.adsabs.harvard.edu/abs/1967JSpRo...4..822B} {4, 822}

\bibitem[\protect\citeauthoryear{{Blondin} \& {Raymer}}{{Blondin} \&
  {Raymer}}{2012}]{2012ApJ...752...30B}
{Blondin} J.~M.,  {Raymer} E.,  2012, \mn@doi [\apj]
  {10.1088/0004-637X/752/1/30}, \href
  {https://ui.adsabs.harvard.edu/abs/2012ApJ...752...30B} {752, 30}

\bibitem[\protect\citeauthoryear{{Cant{\'o}}, {Raga}, {Esquivel}  \&
  {S{\'a}nchez-Salcedo}}{{Cant{\'o}} et~al.}{2011}]{2011MNRAS.418.1238C}
{Cant{\'o}} J.,  {Raga} A.~C.,  {Esquivel} A.,   {S{\'a}nchez-Salcedo} F.~J.,
  2011, \mn@doi [\mnras] {10.1111/j.1365-2966.2011.19574.x}, \href
  {https://ui.adsabs.harvard.edu/abs/2011MNRAS.418.1238C} {418, 1238}

\bibitem[\protect\citeauthoryear{{Chapman}}{{Chapman}}{2000}]{2000hsf..book.....C}
{Chapman} C.~J.,  2000, {High Speed Flow}

\bibitem[\protect\citeauthoryear{{De}, {MacLeod}, {Everson}, {Antoni}, {Mandel}
   \& {Ramirez-Ruiz}}{{De} et~al.}{2020}]{2020ApJ...897..130D}
{De} S.,  {MacLeod} M.,  {Everson} R.~W.,  {Antoni} A.,  {Mandel} I.,
  {Ramirez-Ruiz} E.,  2020, \mn@doi [\apj] {10.3847/1538-4357/ab9ac6}, \href
  {https://ui.adsabs.harvard.edu/abs/2020ApJ...897..130D} {897, 130}

\bibitem[\protect\citeauthoryear{{De} et~al.,}{{De}
  et~al.}{2023}]{engulfmenttransient}
{De} K.,  et~al., 2023, \mn@doi [\nat] {10.1038/s41586-023-05842-x}, 617, 55

\bibitem[\protect\citeauthoryear{{Foglizzo} \& {Ruffert}}{{Foglizzo} \&
  {Ruffert}}{1999}]{1999A&A...347..901F}
{Foglizzo} T.,  {Ruffert} M.,  1999, \aap, \href
  {https://ui.adsabs.harvard.edu/abs/1999A&A...347..901F} {347, 901}

\bibitem[\protect\citeauthoryear{{Foglizzo}, {Galletti}  \&
  {Ruffert}}{{Foglizzo} et~al.}{2005}]{2005A&A...435..397F}
{Foglizzo} T.,  {Galletti} P.,   {Ruffert} M.,  2005, \mn@doi [\aap]
  {10.1051/0004-6361:20042201}, \href
  {https://ui.adsabs.harvard.edu/abs/2005A&A...435..397F} {435, 397}

\bibitem[\protect\citeauthoryear{{Grishin} \& {Perets}}{{Grishin} \&
  {Perets}}{2015}]{2015ApJ...811...54G}
{Grishin} E.,  {Perets} H.~B.,  2015, \mn@doi [\apj]
  {10.1088/0004-637X/811/1/54}, \href
  {https://ui.adsabs.harvard.edu/abs/2015ApJ...811...54G} {811, 54}

\bibitem[\protect\citeauthoryear{{Gurevich}, {Bear}  \& {Soker}}{{Gurevich}
  et~al.}{2022}]{2022MNRAS.511.1330G}
{Gurevich} O.,  {Bear} E.,   {Soker} N.,  2022, \mn@doi [\mnras]
  {10.1093/mnras/stac081}, \href
  {https://ui.adsabs.harvard.edu/abs/2022MNRAS.511.1330G} {511, 1330}

\bibitem[\protect\citeauthoryear{{Hunt}}{{Hunt}}{1971}]{1971MNRAS.154..141H}
{Hunt} R.,  1971, \mn@doi [\mnras] {10.1093/mnras/154.2.141}, \href
  {https://ui.adsabs.harvard.edu/abs/1971MNRAS.154..141H} {154, 141}

\bibitem[\protect\citeauthoryear{Hunter}{Hunter}{2007}]{Hunter:2007}
Hunter J.~D.,  2007, \mn@doi [Computing in Science \& Engineering]
  {10.1109/MCSE.2007.55}, 9, 90

\bibitem[\protect\citeauthoryear{{Jermyn} et~al.,}{{Jermyn}
  et~al.}{2023}]{2023ApJS..265...15J}
{Jermyn} A.~S.,  et~al., 2023, \mn@doi [\apjs] {10.3847/1538-4365/acae8d},
  \href {https://ui.adsabs.harvard.edu/abs/2023ApJS..265...15J} {265, 15}

\bibitem[\protect\citeauthoryear{{Kim} \& {Kim}}{{Kim} \&
  {Kim}}{2009}]{2009ApJ...703.1278K}
{Kim} H.,  {Kim} W.-T.,  2009, \mn@doi [\apj] {10.1088/0004-637X/703/2/1278},
  \href {https://ui.adsabs.harvard.edu/abs/2009ApJ...703.1278K} {703, 1278}

\bibitem[\protect\citeauthoryear{{Kotova}, {Verigin}, {Gombosi}, {Kabin},
  {Slavin}  \& {Bezrukikh}}{{Kotova} et~al.}{2021}]{2021JGRA..12629104K}
{Kotova} G.,  {Verigin} M.,  {Gombosi} T.,  {Kabin} K.,  {Slavin} J.,
  {Bezrukikh} V.,  2021, \mn@doi [Journal of Geophysical Research (Space
  Physics)] {10.1029/2021JA029104}, \href
  {https://ui.adsabs.harvard.edu/abs/2021JGRA..12629104K} {126, e29104}

\bibitem[\protect\citeauthoryear{{Lau}, {Cantiello}, {Jermyn}, {MacLeod},
  {Mandel}  \& {Price}}{{Lau} et~al.}{2022}]{2022arXiv221015848L}
{Lau} M. Y.~M.,  {Cantiello} M.,  {Jermyn} A.~S.,  {MacLeod} M.,  {Mandel} I.,
   {Price} D.~J.,  2022, \mn@doi [arXiv e-prints] {10.48550/arXiv.2210.15848},
  \href {https://ui.adsabs.harvard.edu/abs/2022arXiv221015848L} {p.
  arXiv:2210.15848}

\bibitem[\protect\citeauthoryear{{Livio} \& {Soker}}{{Livio} \&
  {Soker}}{1984}]{1984MNRAS.208..763L}
{Livio} M.,  {Soker} N.,  1984, \mn@doi [\mnras] {10.1093/mnras/208.4.763},
  \href {https://ui.adsabs.harvard.edu/abs/1984MNRAS.208..763L} {208, 763}

\bibitem[\protect\citeauthoryear{Loth}{Loth}{2008}]{doi:10.2514/1.28943}
Loth E.,  2008, \mn@doi [AIAA Journal] {10.2514/1.28943}, 46, 2219

\bibitem[\protect\citeauthoryear{{MacLeod} \& {Ramirez-Ruiz}}{{MacLeod} \&
  {Ramirez-Ruiz}}{2015}]{2015ApJ...803...41M}
{MacLeod} M.,  {Ramirez-Ruiz} E.,  2015, \mn@doi [\apj]
  {10.1088/0004-637X/803/1/41}, \href
  {https://ui.adsabs.harvard.edu/abs/2015ApJ...803...41M} {803, 41}

\bibitem[\protect\citeauthoryear{{MacLeod}, {Antoni}, {Murguia-Berthier},
  {Macias}  \& {Ramirez-Ruiz}}{{MacLeod} et~al.}{2017}]{2017ApJ...838...56M}
{MacLeod} M.,  {Antoni} A.,  {Murguia-Berthier} A.,  {Macias} P.,
  {Ramirez-Ruiz} E.,  2017, \mn@doi [\apj] {10.3847/1538-4357/aa6117}, \href
  {https://ui.adsabs.harvard.edu/abs/2017ApJ...838...56M} {838, 56}

\bibitem[\protect\citeauthoryear{{MacLeod}, {Cantiello}  \&
  {Soares-Furtado}}{{MacLeod} et~al.}{2018}]{2018ApJ...853L...1M}
{MacLeod} M.,  {Cantiello} M.,   {Soares-Furtado} M.,  2018, \mn@doi [\apjl]
  {10.3847/2041-8213/aaa5fa}, \href
  {https://ui.adsabs.harvard.edu/abs/2018ApJ...853L...1M} {853, L1}

\bibitem[\protect\citeauthoryear{{Mai}, {Desch}, {Kuiper}, {Marleau}  \&
  {Dullemond}}{{Mai} et~al.}{2020}]{2020ApJ...899...54M}
{Mai} C.,  {Desch} S.~J.,  {Kuiper} R.,  {Marleau} G.-D.,   {Dullemond} C.,
  2020, \mn@doi [\apj] {10.3847/1538-4357/aba4a8}, \href
  {https://ui.adsabs.harvard.edu/abs/2020ApJ...899...54M} {899, 54}

\bibitem[\protect\citeauthoryear{{Matsuda}, {Inoue}  \& {Sawada}}{{Matsuda}
  et~al.}{1987}]{1987MNRAS.226..785M}
{Matsuda} T.,  {Inoue} M.,   {Sawada} K.,  1987, \mn@doi [\mnras]
  {10.1093/mnras/226.4.785}, \href
  {https://ui.adsabs.harvard.edu/abs/1987MNRAS.226..785M} {226, 785}

\bibitem[\protect\citeauthoryear{{Matsuda}, {Sekino}, {Shima}  \&
  {Sawada}}{{Matsuda} et~al.}{1989}]{1989MNRAS.236..817M}
{Matsuda} T.,  {Sekino} N.,  {Shima} E.,   {Sawada} K.,  1989, \mn@doi [\mnras]
  {10.1093/mnras/236.4.817}, \href
  {https://ui.adsabs.harvard.edu/abs/1989MNRAS.236..817M} {236, 817}

\bibitem[\protect\citeauthoryear{{Matsuda}, {Sekino}, {Sawada}, {Shima},
  {Livio}, {Anzer}  \& {Boerner}}{{Matsuda} et~al.}{1991}]{1991A&A...248..301M}
{Matsuda} T.,  {Sekino} N.,  {Sawada} K.,  {Shima} E.,  {Livio} M.,  {Anzer}
  U.,   {Boerner} G.,  1991, \aap, \href
  {https://ui.adsabs.harvard.edu/abs/1991A&A...248..301M} {248, 301}

\bibitem[\protect\citeauthoryear{{Matsuda}, {Ishii}, {Sekino}, {Sawada},
  {Shima}, {Livio}  \& {Anzer}}{{Matsuda} et~al.}{1992}]{1992MNRAS.255..183M}
{Matsuda} T.,  {Ishii} T.,  {Sekino} N.,  {Sawada} K.,  {Shima} E.,  {Livio}
  M.,   {Anzer} U.,  1992, \mn@doi [\mnras] {10.1093/mnras/255.2.183}, \href
  {https://ui.adsabs.harvard.edu/abs/1992MNRAS.255..183M} {255, 183}

\bibitem[\protect\citeauthoryear{{Metzger}, {Giannios}  \& {Spiegel}}{{Metzger}
  et~al.}{2012}]{2012MNRAS.425.2778M}
{Metzger} B.~D.,  {Giannios} D.,   {Spiegel} D.~S.,  2012, \mn@doi [\mnras]
  {10.1111/j.1365-2966.2012.21444.x}, \href
  {https://ui.adsabs.harvard.edu/abs/2012MNRAS.425.2778M} {425, 2778}

\bibitem[\protect\citeauthoryear{{Meyer}, {Mackey}, {Langer}, {Gvaramadze},
  {Mignone}, {Izzard}  \& {Kaper}}{{Meyer} et~al.}{2014}]{2014MNRAS.444.2754M}
{Meyer} D.~M.~A.,  {Mackey} J.,  {Langer} N.,  {Gvaramadze} V.~V.,  {Mignone}
  A.,  {Izzard} R.~G.,   {Kaper} L.,  2014, \mn@doi [\mnras]
  {10.1093/mnras/stu1629}, \href
  {https://ui.adsabs.harvard.edu/abs/2014MNRAS.444.2754M} {444, 2754}

\bibitem[\protect\citeauthoryear{{Mignone}}{{Mignone}}{2014}]{2014JCoPh.270..784M}
{Mignone} A.,  2014, \mn@doi [Journal of Computational Physics]
  {10.1016/j.jcp.2014.04.001}, \href
  {https://ui.adsabs.harvard.edu/abs/2014JCoPh.270..784M} {270, 784}

\bibitem[\protect\citeauthoryear{{Morton}, {Khochfar}  \&
  {O{\~n}orbe}}{{Morton} et~al.}{2021}]{2021arXiv210315848M}
{Morton} B.,  {Khochfar} S.,   {O{\~n}orbe} J.,  2021, \mn@doi [arXiv e-prints]
  {10.48550/arXiv.2103.15848}, \href
  {https://ui.adsabs.harvard.edu/abs/2021arXiv210315848M} {p. arXiv:2103.15848}

\bibitem[\protect\citeauthoryear{{Nelemans} \& {Tauris}}{{Nelemans} \&
  {Tauris}}{1998}]{1998A&A...335L..85N}
{Nelemans} G.,  {Tauris} T.~M.,  1998, \mn@doi [\aap]
  {10.48550/arXiv.astro-ph/9806011}, \href
  {https://ui.adsabs.harvard.edu/abs/1998A&A...335L..85N} {335, L85}

\bibitem[\protect\citeauthoryear{{O'Connor}, {Bildsten}, {Cantiello}  \&
  {Lai}}{{O'Connor} et~al.}{2023}]{2023ApJ...950..128O}
{O'Connor} C.~E.,  {Bildsten} L.,  {Cantiello} M.,   {Lai} D.,  2023, \mn@doi
  [\apj] {10.3847/1538-4357/acd2d4}, \href
  {https://ui.adsabs.harvard.edu/abs/2023ApJ...950..128O} {950, 128}

\bibitem[\protect\citeauthoryear{{Ostriker}}{{Ostriker}}{1999}]{1999ApJ...513..252O}
{Ostriker} E.~C.,  1999, \mn@doi [\apj] {10.1086/306858}, \href
  {https://ui.adsabs.harvard.edu/abs/1999ApJ...513..252O} {513, 252}

\bibitem[\protect\citeauthoryear{{Paxton}, {Bildsten}, {Dotter}, {Herwig},
  {Lesaffre}  \& {Timmes}}{{Paxton} et~al.}{2011}]{2011ApJS..192....3P}
{Paxton} B.,  {Bildsten} L.,  {Dotter} A.,  {Herwig} F.,  {Lesaffre} P.,
  {Timmes} F.,  2011, \mn@doi [\apjs] {10.1088/0067-0049/192/1/3}, \href
  {http://adsabs.harvard.edu/abs/2011ApJS..192....3P} {192, 3}

\bibitem[\protect\citeauthoryear{{Paxton} et~al.,}{{Paxton}
  et~al.}{2013}]{2013ApJS..208....4P}
{Paxton} B.,  et~al., 2013, \mn@doi [\apjs] {10.1088/0067-0049/208/1/4}, \href
  {http://adsabs.harvard.edu/abs/2013ApJS..208....4P} {208, 4}

\bibitem[\protect\citeauthoryear{{Paxton} et~al.,}{{Paxton}
  et~al.}{2015}]{2015ApJS..220...15P}
{Paxton} B.,  et~al., 2015, \mn@doi [\apjs] {10.1088/0067-0049/220/1/15}, \href
  {http://adsabs.harvard.edu/abs/2015ApJS..220...15P} {220, 15}

\bibitem[\protect\citeauthoryear{{Paxton} et~al.,}{{Paxton}
  et~al.}{2018}]{2018ApJS..234...34P}
{Paxton} B.,  et~al., 2018, \mn@doi [\apjs] {10.3847/1538-4365/aaa5a8}, \href
  {https://ui.adsabs.harvard.edu/abs/2018ApJS..234...34P} {234, 34}

\bibitem[\protect\citeauthoryear{{Paxton} et~al.,}{{Paxton}
  et~al.}{2019}]{2019ApJS..243...10P}
{Paxton} B.,  et~al., 2019, \mn@doi [\apjs] {10.3847/1538-4365/ab2241}, \href
  {https://ui.adsabs.harvard.edu/abs/2019ApJS..243...10P} {243, 10}

\bibitem[\protect\citeauthoryear{{Rephaeli} \& {Salpeter}}{{Rephaeli} \&
  {Salpeter}}{1980}]{1980ApJ...240...20R}
{Rephaeli} Y.,  {Salpeter} E.~E.,  1980, \mn@doi [\apj] {10.1086/158202}, \href
  {https://ui.adsabs.harvard.edu/abs/1980ApJ...240...20R} {240, 20}

\bibitem[\protect\citeauthoryear{{Ruderman} \& {Spiegel}}{{Ruderman} \&
  {Spiegel}}{1971}]{1971ApJ...165....1R}
{Ruderman} M.~A.,  {Spiegel} E.~A.,  1971, \mn@doi [\apj] {10.1086/150870},
  \href {https://ui.adsabs.harvard.edu/abs/1971ApJ...165....1R} {165, 1}

\bibitem[\protect\citeauthoryear{{Ruffert}}{{Ruffert}}{1994}]{1994ApJ...427..342R}
{Ruffert} M.,  1994, \mn@doi [\apj] {10.1086/174144}, \href
  {https://ui.adsabs.harvard.edu/abs/1994ApJ...427..342R} {427, 342}

\bibitem[\protect\citeauthoryear{{Ruffert} \& {Arnett}}{{Ruffert} \&
  {Arnett}}{1994}]{1994ApJ...427..351R}
{Ruffert} M.,  {Arnett} D.,  1994, \mn@doi [\apj] {10.1086/174145}, \href
  {http://adsabs.harvard.edu/abs/1994ApJ...427..351R} {427, 351}

\bibitem[\protect\citeauthoryear{{S{\'a}nchez-Salcedo} \&
  {Brandenburg}}{{S{\'a}nchez-Salcedo} \&
  {Brandenburg}}{1999}]{1999ApJ...522L..35S}
{S{\'a}nchez-Salcedo} F.~J.,  {Brandenburg} A.,  1999, \mn@doi [\apjl]
  {10.1086/312215}, \href
  {https://ui.adsabs.harvard.edu/abs/1999ApJ...522L..35S} {522, L35}

\bibitem[\protect\citeauthoryear{{S{\'a}nchez-Salcedo} \&
  {Brandenburg}}{{S{\'a}nchez-Salcedo} \&
  {Brandenburg}}{2001}]{2001MNRAS.322...67S}
{S{\'a}nchez-Salcedo} F.~J.,  {Brandenburg} A.,  2001, \mn@doi [\mnras]
  {10.1046/j.1365-8711.2001.04061.x}, \href
  {https://ui.adsabs.harvard.edu/abs/2001MNRAS.322...67S} {322, 67}

\bibitem[\protect\citeauthoryear{{Scherer}, {Fichtner}, {Kleimann},
  {Wiengarten}, {Bomans}  \& {Weis}}{{Scherer}
  et~al.}{2016}]{2016A&A...586A.111S}
{Scherer} K.,  {Fichtner} H.,  {Kleimann} J.,  {Wiengarten} T.,  {Bomans}
  D.~J.,   {Weis} K.,  2016, \mn@doi [\aap] {10.1051/0004-6361/201526137},
  \href {https://ui.adsabs.harvard.edu/abs/2016A&A...586A.111S} {586, A111}

\bibitem[\protect\citeauthoryear{{Schneider}, {Dedieu}, {Le Sidaner}, {Savalle}
   \& {Zolotukhin}}{{Schneider} et~al.}{2011}]{2011A&A...532A..79S}
{Schneider} J.,  {Dedieu} C.,  {Le Sidaner} P.,  {Savalle} R.,   {Zolotukhin}
  I.,  2011, \mn@doi [\aap] {10.1051/0004-6361/201116713}, \href
  {https://ui.adsabs.harvard.edu/abs/2011A&A...532A..79S} {532, A79}

\bibitem[\protect\citeauthoryear{{Shara} \& {Shaviv}}{{Shara} \&
  {Shaviv}}{1980}]{1980Ap&SS..67..427S}
{Shara} M.~M.,  {Shaviv} G.,  1980, \mn@doi [\apss] {10.1007/BF00642396}, \href
  {https://ui.adsabs.harvard.edu/abs/1980Ap&SS..67..427S} {67, 427}

\bibitem[\protect\citeauthoryear{{Shima}, {Matsuda}, {Takeda}  \&
  {Sawada}}{{Shima} et~al.}{1985}]{1985MNRAS.217..367S}
{Shima} E.,  {Matsuda} T.,  {Takeda} H.,   {Sawada} K.,  1985, \mn@doi [\mnras]
  {10.1093/mnras/217.2.367}, \href
  {https://ui.adsabs.harvard.edu/abs/1985MNRAS.217..367S} {217, 367}

\bibitem[\protect\citeauthoryear{{Shima}, {Matsuda}, {Anzer}, {Boerner}  \&
  {Boffin}}{{Shima} et~al.}{1998}]{1998A&A...337..311S}
{Shima} E.,  {Matsuda} T.,  {Anzer} U.,  {Boerner} G.,   {Boffin} H. M.~J.,
  1998, \mn@doi [\aap] {10.48550/arXiv.astro-ph/9805343}, \href
  {https://ui.adsabs.harvard.edu/abs/1998A&A...337..311S} {337, 311}

\bibitem[\protect\citeauthoryear{Stone, Tomida, White  \& Felker}{Stone
  et~al.}{2020}]{athena++}
Stone J.~M.,  Tomida K.,  White C.~J.,   Felker K.~G.,  2020, \mn@doi [The
  Astrophysical Journal Supplement Series] {10.3847/1538-4365/ab929b}, 249, 4

\bibitem[\protect\citeauthoryear{{Sz{\"o}lgy{\'e}n}, {MacLeod}  \&
  {Loeb}}{{Sz{\"o}lgy{\'e}n} et~al.}{2022}]{2022MNRAS.513.5465S}
{Sz{\"o}lgy{\'e}n} {\'A}.,  {MacLeod} M.,   {Loeb} A.,  2022, \mn@doi [\mnras]
  {10.1093/mnras/stac1294}, \href
  {https://ui.adsabs.harvard.edu/abs/2022MNRAS.513.5465S} {513, 5465}

\bibitem[\protect\citeauthoryear{{Tarango-Yong} \& {Henney}}{{Tarango-Yong} \&
  {Henney}}{2018}]{2018MNRAS.477.2431T}
{Tarango-Yong} J.~A.,  {Henney} W.~J.,  2018, \mn@doi [\mnras]
  {10.1093/mnras/sty669}, \href
  {https://ui.adsabs.harvard.edu/abs/2018MNRAS.477.2431T} {477, 2431}

\bibitem[\protect\citeauthoryear{{Thun}, {Kuiper}, {Schmidt}  \& {Kley}}{{Thun}
  et~al.}{2016}]{thun}
{Thun} D.,  {Kuiper} R.,  {Schmidt} F.,   {Kley} W.,  2016, \mn@doi [\aap]
  {10.1051/0004-6361/201527629}, \href
  {https://ui.adsabs.harvard.edu/abs/2016A&A...589A..10T} {589, A10}

\bibitem[\protect\citeauthoryear{{Toro}, {Spruce}  \& {Speares}}{{Toro}
  et~al.}{1994}]{1994ShWav...4...25T}
{Toro} E.~F.,  {Spruce} M.,   {Speares} W.,  1994, \mn@doi [Shock Waves]
  {10.1007/BF01414629}, \href
  {https://ui.adsabs.harvard.edu/abs/1994ShWav...4...25T} {4, 25}

\bibitem[\protect\citeauthoryear{{Villaver}, {Livio}, {Mustill}  \&
  {Siess}}{{Villaver} et~al.}{2014}]{2014ApJ...794....3V}
{Villaver} E.,  {Livio} M.,  {Mustill} A.~J.,   {Siess} L.,  2014, \mn@doi
  [\apj] {10.1088/0004-637X/794/1/3}, \href
  {https://ui.adsabs.harvard.edu/abs/2014ApJ...794....3V} {794, 3}

\bibitem[\protect\citeauthoryear{Virtanen et~al.,}{Virtanen
  et~al.}{2020}]{2020SciPy-NMeth}
Virtanen P.,  et~al., 2020, \mn@doi [Nature Methods]
  {10.1038/s41592-019-0686-2}, \href {https://rdcu.be/b08Wh} {17, 261}

\bibitem[\protect\citeauthoryear{{Wright} et~al.,}{{Wright}
  et~al.}{2011}]{2011PASP..123..412W}
{Wright} J.~T.,  et~al., 2011, \mn@doi [\pasp] {10.1086/659427}, \href
  {https://ui.adsabs.harvard.edu/abs/2011PASP..123..412W} {123, 412}

\bibitem[\protect\citeauthoryear{{Xu} \& {Stone}}{{Xu} \&
  {Stone}}{2019}]{2019MNRAS.488.5162X}
{Xu} W.,  {Stone} J.~M.,  2019, \mn@doi [\mnras] {10.1093/mnras/stz2002}, \href
  {https://ui.adsabs.harvard.edu/abs/2019MNRAS.488.5162X} {488, 5162}

\bibitem[\protect\citeauthoryear{{Yarza} et~al.,}{{Yarza} et~al.}{2022}]{yarza}
{Yarza} R.,  et~al., 2022, \mn@doi [arXiv e-prints]
  {10.48550/arXiv.2203.11227}, \href
  {https://ui.adsabs.harvard.edu/abs/2022arXiv220311227Y} {p. arXiv:2203.11227}

\makeatother
\end{thebibliography}


\appendix
\section{Convergence Tests} \label{sec:convergence}

\begin{figure}
  \includegraphics[width=0.5\textwidth]{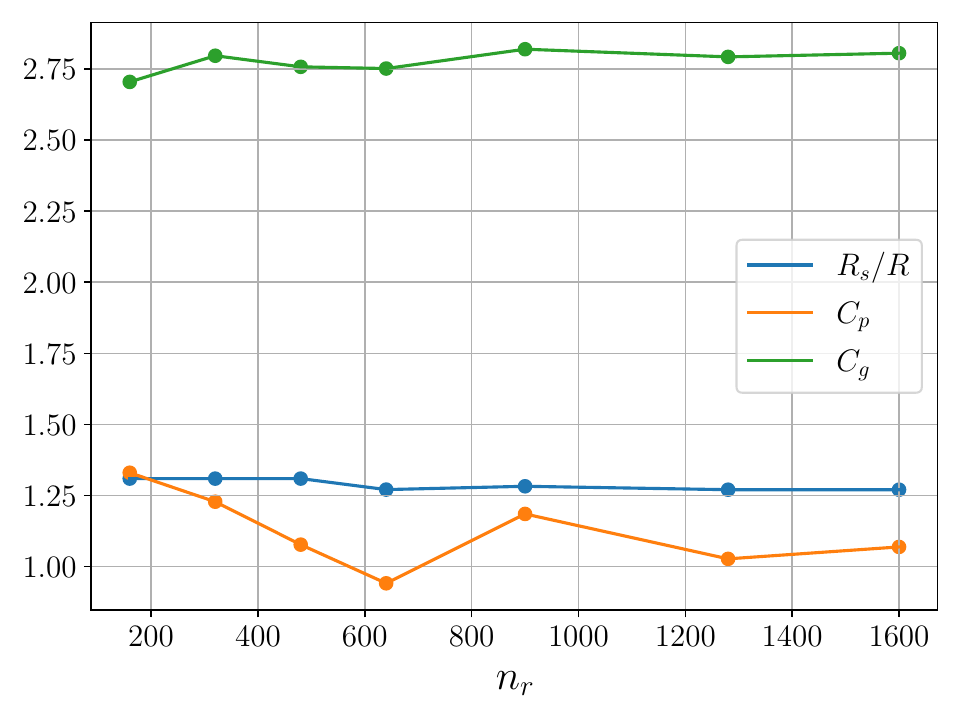}
    \caption{Convergence of several of our key derived quantities with linear resolution. The stand-off distance and gravitational drag coefficient converge nicely, while the pressure drag exhibits some fluctuations even at high resolution.
    \label{fig:conv}}
\end{figure}

We check that our results are converged by varying the linear resolution of our grid. Setting the size of the domain at $R_{\rm out}=100R$, we vary the number of cells in the radial direction $n_{r}$. The cells in the polar and azimuthal directions $n_{\theta}$ and $n_{\phi}$ are scaled proportionally. We show in Fig. \ref{fig:conv} the change in our principal derived quantities $\rs$, $C_{p}$  and $C_{g}$ as a function of $n_{r}$. We see that for $n_{r}\geq640$ the shock stand-off distance and gravitational drag coefficient vary on the percent level. The coefficient of pressure drag exhibits significant fluctuations, though these do not disappear even at high resolution. We conclude that our results for $\rs$ and $C_{g}$ are converged, while the error in $C_{p}$ may be of order $\approx$10 per cent.

\begin{figure}
  \includegraphics[width=0.5\textwidth]{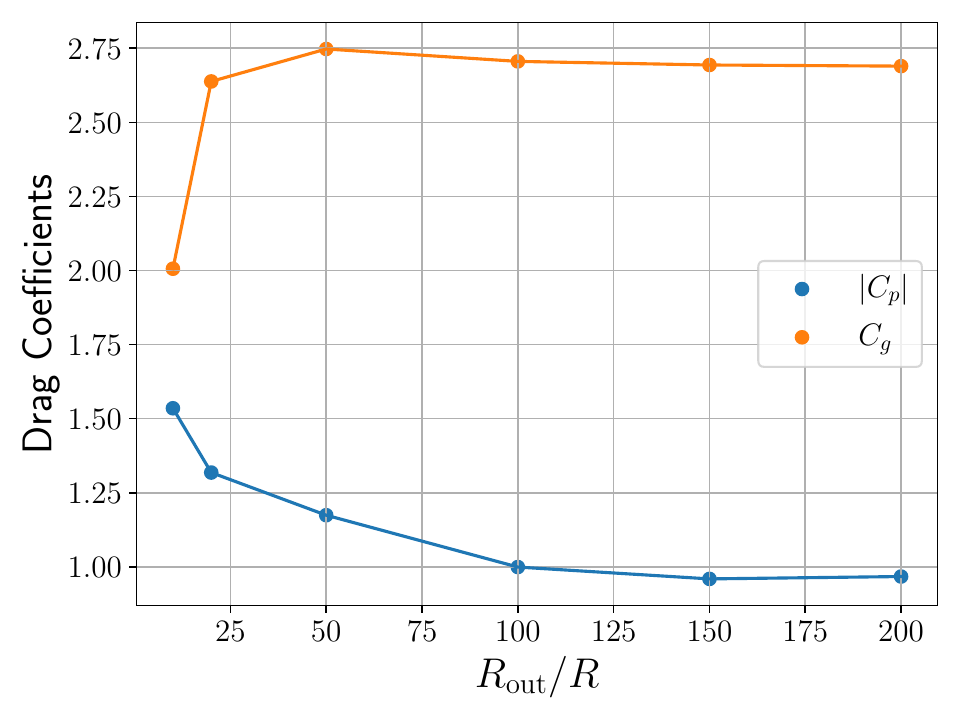}
    \caption{Convergence of drag coefficients with the domain radius $R_{\rm out}$, fixing $\eta=1$ and $\mach=4$. Both coefficients vary by $\sim$50 per cent, illustrating the dangers of using a domain of insufficient size.
    \label{fig:boxsize}}
\end{figure}

We have assumed in this work that the streamlines of the fluid entering the domain from the boundary are initially parallel. However, in the presence of gravity the streamlines should have deflected by some amount even at arbitrarily large $r$. One solution is to approximate the deflection angle of the streamlines at some radius from the gravitating body, though it is not possible to write an exact solution for this deflection. We choose to tackle this issue by ensuring that the radius of our domain is large enough that the deflection at the outer boundary is negligible. To this end, we perform a series of runs with $\eta=1$ and $\mach=4$ with varying values of $R_{\rm out}$, as shown in Table \ref{tab:setups}. As the domain is enlarged, we increase the number of cells in the radial direction such that the linear resolution is preserved. Thus, any discrepancies between these simulations can be attributed purely to the domain size. Our results in Fig. \ref{fig:boxsize} show that the drag coefficients vary by $\sim$50 per cent before convergence, illustrating the large errors that can result from a domain of insufficient size. The flow morphology is also affected by the streamline deflection. From Table \ref{tab:setups} we see that the extent of the separation bubble $\theta_{b}$ changes by over 12 degrees as the domain size is varied, even cresting the top of the sphere ($\theta=90$ deg). We do not find any change in the shock stand-off distance. The majority of the calculations presented in this paper use $R_{\rm out}=100R$, which appears to be near convergence for both $C_{p}$ and $C_{g}$. For high $\eta$ values, we extend the domain size to accommodate the large accretion radii, scaling by $\eta$ such that $R_{\rm out}=100\eta R \approx 50\ra$. The refinement boundaries are also scaled by $\eta$.

\section{Halo Mass} \label{sec:halomass}

The density profile in the hydrostatic halo, as shown in T16 and confirmed in section \ref{sec:halo}, is
\be
\rho(r) = \rho_{2} \left[1-\frac{\gamma-1}{c_{s,2}^{2}/c_{s,\infty}^{2}}\frac{\mach_{\infty}^{2}}{2}\left(\frac{\ra}{\rs}-\frac{\ra}{r}\right)\right]^{1/(\gamma-1)}. \label{eq:rhor}
\ee
Here the subscript 2 denotes post-shock conditions. Conservation of energy gives the pre-shock Mach number
\be
\mach_{1}^{2} = \mach_{\infty}^{2} \left(1+\frac{\ra}{\rs}\right),
\ee
and because $\rs=\eta R$ at high $\eta$, this simplifies to 
\be
\mach_{1}^{2} = 3\mach_{\infty}^{2}-2.
\ee
This can then be used in conjunction with the jump conditions and ambient medium parameters to find the post-shock fluid state. Using the definitions of $\ra$ and $h$, we rewrite (\ref{eq:rhor}) as
\be
\rho(r) = \rho_{2} \left[1-\frac{G m}{h_{2}}\left(\frac{1}{\rs}-\frac{1}{r}\right)\right]^{1/(\gamma-1)}.
\ee
Then the total mass contained within the halo out to $r=\rs$ is
\be
\begin{aligned}
m_{\rm halo} &= \int \rho(r) d^{3}r \\
&= 4\pi\rho_{2}\left(\frac{G m}{h_{2}}\right)^{1/(\gamma-1)} \int_{R}^{\rs} r^{2} \left( \chi + \frac{1}{r} \right)^{1/(\gamma-1)} dr,
\end{aligned}
\ee
where we have introduced the variable
\be
\chi=\frac{h_{2}}{Gm}-\frac{1}{\rs}.
\ee
Solutions to integrals of this form involve the hypergeometric function
\be
_{2}F_{1}\left(4,\frac{\gamma}{\gamma-1},\frac{2\gamma-1}{\gamma-1},1+\frac{1}{\chi r}\right),
\ee
which for the special case of $\gamma=5/3$ reduces to the exact solution
\be
\begin{aligned}
m_{\rm halo} = \pi \rho_{2} \left( \frac{G m}{h_{2} \chi} \right)^{3/2} \left[\chi \rs\sqrt{\frac{1}{\chi \rs}+1}\left(\frac{4}{3}\chi^{2}\rs^{2}+\frac{7}{3}\chi\rs+\frac{1}{2}\right)\right. \\ \left.-\frac{1}{2}\arctanh\sqrt{\frac{1}{\chi\rs}+1} - \chi R\sqrt{\frac{1}{\chi R}+1}\left(\frac{4}{3}\chi^{2}R^{2}+\frac{7}{3}\chi R+\frac{1}{2}\right)\right. \\ \left.+\frac{1}{2}\arctanh\sqrt{\frac{1}{\chi R}+1}\right]. \label{eq:big}
\end{aligned}
\ee
Fortunately, we can make several simplifications about the terms within the brackets in the above equation:
\begin{enumerate}[(i)]
    \item The quantity $\chi\rs$ can be rewritten as
    \be
    \chi\rs = \frac{h_{2}}{G m} - 1 = \frac{h_{2}}{v_{\infty}^{2}}\frac{\mach^{2}}{\mach^{2}-1}-1, \label{eq:chirs}
    \ee
    which depends only on the upstream conditions. Thus, terms depending only on $\chi\rs$ (first two terms) do not scale with $\eta$. \\

    \item Similarly, we can write
    \be
    \chi R = \frac{1}{\eta}\left(\frac{h_{2}}{v_{\infty}^{2}}\frac{\mach^{2}}{\mach^{2}-1}-1\right). \label{eq:chir}
    \ee
    For large $\eta$, this is negligible compared to (\ref{eq:chirs}), so we can ignore terms which scale with $\chi R$ (third term). This is equivalent to the assumption $\rs\gg R$. \\

    \item Using (\ref{eq:chir}), we can compute the derivative of the final term with respect to $\eta$:
    \be
    \frac{\partial}{\partial\eta}\left(\frac{1}{2}\arctanh\sqrt{\frac{1}{\chi R}+1}\right) = \frac{-1}{4\eta\sqrt{\eta(\frac{h_{2}}{v_{\infty}^{2}}\frac{\mach^{2}}{\mach^{2}-1}-1)^{-1}+1}}.
    \ee
    This derivative also vanishes for $\eta\gg1$, so this term depends only weakly on $\eta$ in a hydrostatic halo.
\end{enumerate}
From the discussion above, we conclude that the quantity in brackets in (\ref{eq:big}) does not scale significantly with $\eta$. Thus, the scaling is determined by the prefactor:
\be
m_{\rm halo} \propto \rho_{2}\left(\frac{Gm}{h_{2}\chi}\right)^{3/2}.
\ee
Furthermore, since the jump conditions give
\be
\rho_{2} = \rho_{1}\frac{(\gamma+1)\mach_{1}^{2}}{(\gamma-1)\mach_{1}^{2}+2} = \rho_{\infty}\frac{(\gamma+1)(3\mach_{\infty}^{2}-2)}{(\gamma-1)(3\mach_{\infty}^{2}-2)+2},
\ee
we can plug $\chi$ back in to obtain
\be
m_{\rm halo} \propto \rho_{\infty}\left(\frac{Gm}{h_{2}\chi}\right)^{3/2} = \rho_{\infty}\left(\frac{h_{2}^{2}}{G^{2}m^{2}}-\frac{h_{2}}{Gm\eta R}\right)^{-3/2}.
\ee
Using the definitions of $\ra$ and $\eta$, this becomes
\be
m_{\rm halo} \propto \rho_{\infty}\left(\frac{1}{4\ra^{2}}\frac{h_{2}}{v_{\infty}^{2}}\left[\frac{h_{2}}{v_{\infty}^{2}}-\cos^{2}\mu\right]\right)^{-3/2},
\ee
which scales as
\be
m_{\rm halo} \propto \rho_{\infty}\left(\frac{1}{\ra^{2}}\right)^{-3/2},
\ee
or
\be
m_{\rm halo} \propto \rho_{\infty}\ra^{3}.
\ee
Therefore, for given upstream flow conditions the mass of the hydrostatic halo within $\rs$ scales as the ambient density times the cube of the accretion radius, or equivalently as $\rho_{\infty}\eta^{3}$.

\bsp	
\label{lastpage}
\end{document}